\documentclass[useAMS,usenatbib,onecolumn]{mn2e}

\usepackage{graphicx}
\title[Magnetic Jets from Swirling Disks]{Magnetic Jets from Swirling Disks}
\author[D. Lynden-Bell]{D. Lynden-Bell$^{1,2}$\\
$^{1}$Institute of Astronomy, The Observatories, Cambridge, CB3 0HA\\
$^{2}$Clare College}

\begin{document}

\date{This version dated 24 January 2006, revised 3 February 2006,
revised following reviewer's comments 20 March 2006}

\pagerange{\pageref{firstpage}--\pageref{lastpage}} \pubyear{2006}

\maketitle

\label{firstpage}

\begin{abstract}
A broad swathe of astrophysical phenomena, ranging from tubular
planetary nebulae through Herbig-Haro objects, radio-galaxy and quasar
emissions to gamma-ray bursts and perhaps high-energy cosmic rays, may
be driven by magnetically-dominated jets emanating from accretion
disks. We give a self-contained account of the analytic theory of
non-relativistic magnetically dominated jets wound up by a swirling
disk and making a magnetic cavity in a background medium of any
prescribed pressure, $p(z)$. We solve the time-dependent problem for
any specified distribution of magnetic flux $P(R,0)$ emerging from the
disk at $z=0$, with any specified disk angular velocity
$\Omega_d(R)$. The physics required to do this involves only the
freezing of the lines of force to the conducting medium and the
principle of minimum energy.

In a constant pressure environment the magnetically dominated cavity
is highly collimated \/ and advances along the axis \/ at a constant
speed closely related to the maximum circular velocity of the
accretion disk. Even within the cavity the field is strongly
concentrated toward the axis. The twist in the jet's field
$<B_{\phi}>/<|B_z|>$\\ is close to $\sqrt2$ and the width of the jet
decreases upwards. By contrast when the background pressure falls off
with height with powers approaching $z^{-4}$ the head of the jet
accelerates strongly and the twist of the jet is much smaller. The
width increases to give an almost conical magnetic cavity with apex at
the source.  Such a regime may be responsible for some of the longest
strongly collimated jets. When the background pressure falls off
faster than $z^{-4}$ there are no quasi-static configurations of well
twisted fields and the pressure confinement is replaced by a dynamic
effective pressure or a relativistic expansion. In the regimes with
rapid acceleration the outgoing and incoming fields linking the twist
back to the source are almost anti-parallel so there is a possibility
that magnetic reconnections may break up the jet into a series of
magnetic `smoke-rings' travelling out along the axis.
\end{abstract}

\begin{keywords}
jets, MHD, Cosmic Rays
\end{keywords}

\section{Introduction}
\subsection{Orders of magnitude -- the voltages generated}
We consider accretion disks of bodies in formation when the
differential rotation drags around magnetic field lines. Although the
moving magnetic fields inevitably generate electric fields, the
resulting EMFs are perpendicular to the magnetic fields in the perfect
conduction approximation.  Such EMFs do not accelerate particles to
high energies. However the world is not a perfect place; in regions
where perfect MHD predicts very high current densities there may be
too few charge carriers to carry those currents. In those regions
perfect conductivity is not a good approximation and the fields are
modified to allow an electric field component along the magnetic field
so that the larger current can be generated by the few charge carriers
available.  With this background idea it becomes interesting to get
rough estimates of EMFs that are likely to be around whatever their
direction. It turns out that these EMFs scale with $(v/c)^3$ where $v$
is the maximum velocity in the accretion disk and is independent of
the size or mass of the system.  Relativistic systems in formation,
however small, can generate EMFs whose voltages at least match those
of the highest energy cosmic rays. However the timescales over which
these systems persist and the total energies available over those time
scales do of course depend on the mass of the system.  Very crude
estimates of the EMFs can be made as follows. Let us suppose that
there is a central object, a star or a black hole, of mass $M$
surrounded by an accretion disk of mass $\zeta M$. Further let us
suppose that a small fraction $\eta$ of the binding energy of the
accretion disk is converted into magnetic field energy. Letting R be
the inner radius of the disk we put
$(4\pi/3)R^3(B^2/(8\pi))=G\zeta\eta M^2/R$ so
$B=(6\zeta\eta/G)^{1/2}v^2/R$., where $v^2=GM/R$. Fields of a few
hundred Gauss are found in T Tauri stars, so putting $R=10^{11}$cm
$v=300$km/s we find $\zeta \eta=10^{-11}$. We shall use this
dimensionless number to make estimates elsewhere. Now the EMF
generated around a circuit that goes up the axis and back to the disk
around a field line and finally along the disk back to the centre is
about $(v/c)BR$ in esu and 300 times that in volts so EMF= 300
$(6\zeta \eta/G)^{1/2}(v/c)^3 c^2=(v/c)^3 10^{22}$ volts. Kronberg et
al (2004) considered such accelerative processes in radio galaxies.

Again only a fraction of this voltage will be available to accelerate
particles; in the exact flux freezing case it is ALL perpendicular to the
fields, nevertheless it is an estimate of what might be available where the
perfect conductivity approximation breaks down. Although the argument is
crude the answer is interesting in that it suggests that cosmic rays generated
in microquasars may reach the same individual particle energies as those
generated in quasars and radio galaxies which may match the highest cosmic ray
energies.

\subsection{Jets}
Curtis photographed the jet in M 87 from the Lick Observatory in
1918. While it was soon found to be blue, the emission process was not
understood despite Schott's detailed calculations of synchrotron
radiation in his 1912 book. Finally after Shklovskii introduced this
emission mechanism into astrophysics Baade (1956) showed the jet to be
highly polarized which clinched it. Although Ryle's Cambridge group
(1968) found many double radio lobes around large galaxies the next
obvious jet came with the identification of the first quasar 3C273 in
which the dominant radio source is not at the nucleus but at the other
end of the optical jet (Hazard, Mackey \& Shimmins 1963; Schmidt
1963). There were difficulties in understanding the powering of the
radio lobes of galaxies. If all the energy were present as the lobes
expanded outwards there would be more bright small ones. Finally this
led Rees (1971) to suggest that the lobes must be continuously powered
by as-yet-unseen jets feeding energy into the visible lobes. As radio
astronomy moved to higher frequencies with greater resolution these
jets duly appeared in both radio galaxies and quasars. All the above
jets have dynamically significant magnetic fields seen via their
synchrotron radiation; a particularly fine study of one, Herculis A,
is found in Gizani \& Leahy (2003). Magnetic fields are less obvious
in the Herbig-Haro objects first found in star-forming regions in
1951. However it took the development of good infra-red detectors
before the heavy obscuration was penetrated in the 1980s to reveal the
jets feeding the emission. These jets have velocities of one or two
hundred km/s, far less than the $0.1c - c$ speeds of the extragalactic
jets. The jets around young stars are seen to be perpendicular to the
accretion discs that generate them. Since the giant black-hole
accretion disk theory of quasars Salpeter (1964), Lynden-Bell (1969),
Bardeen (1970), Lynden-Bell \& Rees (1971) Shakura \& Sunyaev (1973,
1976) many have come to believe that the radio galaxies and quasars
likewise have jets perpendicular to their inner disks. However it was
not until the wonderful work on megamasers (Miyoshi et al. 1995) that
such inner disks around black holes were definitively confirmed. In
1969 when I predicted that they would inhabit the nuclei of most major
galaxies including our own, M31, M32, M81, M87, etc., the idea was
considered outlandish, but now most astronomers take it for
granted. Fine work by Kormendy (1995) and others on external galaxies
and the beautiful results of Genzel (2003) and Ghez (2004) on our own
has totally transformed the situation. Meanwhile many jets have been
found in objects associated with dying stars, SS433 and the
micro-quasars Mirabel \& Rodriguez (1999) being prominent examples
within the Galaxy. Less energetic but more beautiful examples may be
the tubular planetary nebulae that are associated with accretion discs
of central mass-exchanging binary stars. These were brought to my
attention by Mark Morris and recent evidence indicates that magnetism
is important here too (Vlemmings et al 2006). Much more spectacularly
the $\gamma$-ray bursts are now thought to come from accretion-disks
around black holes within some supernova explosions. Poynting flows of
electromagnetic energy appear to be one of the best ways of extracting
the energy from beneath the baryons that would otherwise absorb the
$\gamma$-rays, Uzdensky \& MacFadyen (2006). Jets and collimated
outflows have also been invoked for giant stars. Remarkable examples
are R Aquarii (Michalitsanos et al. 1988) and the Egg nebula (Cohen et
al. 2004) and the collimated outflow in IRG 10011 (Vinkovi\'c 2004).

The systematic features of these diverse objects are that an accretion
disk is present and the jet emerges along the rotation axis of the
inner disk.  Magnetic fields are important in the radio objects and
may be important in all. The jet velocities are strongly correlated
with the escape velocities and therefore with the circular velocities
in the disk close to the central object. In very collapsed objects
these velocities are relativistic but in star-forming regions the jet
velocities are less than $10^{-3}c$. The obvious similarity of the jet
structures and collimation despite such velocity differences strongly
suggest that relativity is not a determining factor in the making of
these jets. The thesis that I put forward in 1996 (paper II) and
repeat here, is that all these jets are magnetically driven, the
common feature being that the Poynting flux dominates the energy
transport in the jet. However I do not exclude the possibility of
material being entrained as the jet makes its way through the material
that surrounds it. Nevertheless we consider that magnetism is the
driver and the prime reason for collimation is the magnetic twisting
combined with a weak external pressure which is dynamically enhanced
by inertia. It is the electromagnetic field that generates the
relativistic particles in radio jets so, even if their total energies
evolve to become comparable, the magnetic energy comes first and
drives the whole phenomenon. While this paper is concerned with jets
from systems with accretion disks, the jets near pulsars are probably
closely related. There the magnetic field is rotated by the neutron
star but the weaker field at large radii may be heavily loaded by the
inertia close to the light cylinder. Field lines may be twisted from
below far faster than they can rotate across the light cylinder. The
resulting twisting up of the field within the cylinder may result in
jet phenomena with similarities to those described here.

The problem of jet collimation was emphasised by Wheeler (1971) at the
Vatican Conference on the Nuclei of Galaxies in 1970. He drew
attention to the computations of Leblanc \& Wilson (1970) who found a
remarkable jet generated on the axis of a rotating star in
collapse. This may be hailed as the first gamma-ray burst calculation
and the magnetic cavities discussed here are in essence analytic
calculations based on the same mechanism. We have concentrated on the
simplest case of force-free magnetic fields within the cavity. In
earlier work others produced good collimation by twisting up a
magnetic field that was imposed from infinity. An early paper on this
was by Lovelace (1976) and the rather successful simulations of
Shibata \& Uchida (1985, 1986) are based on this theme. I consider
that the imposition of a straight field from infinity ducks the
question of why collimation exists.  The straight field imposes
it. Much work has concentrated on the hard problem of winds carrying a
significant mass flux from a rotating star. This subject is well
covered in Mestel's book (1999) Li et al (2001), Lovelace et al (2002)
and Sakarai (1987) found a weak asymptotic collimation and Heyvaerts
\& Norman (2002) have recently concluded a thorough study of the
asymptotics of wind collimation. Bogovalov \& Tsinganos (2003) have
tackled the difficulties of collimating mass loaded flows from central
objects and give models with the near-axis part of the flow well
collimated despite the centrifugal force. Blandford \& Payne (1982)
discuss winds launched by centrifugal force. Lovelace \& Romanova
(2003) Li et al (2006) made numerical calculations based upon the
differential winding we proposed in Papers I \& II. The stability
problems of twisted jets have been tackled in both the linear and
non-linear regimes by Appl, Lery \& Baty (2000) and Lery, Baty \& Appl
(2000) while Thompson, Lyutikov and Kulkarni (2002) have applied to
magnetars the self-similar fields found in Paper I.  In section 5 we
find that some magnetic cavities float upwards like bubbles thus
fulfilling the ideas of Gull \& Northover (1976). The laboratory
experiments of Lebedev et al. (2005) give some support for the type of
models given here. A fine review of extragalactic jets was given by
Begelman et al. (1984). See also the more recent work of Pudritz et al
2006, Ouyed et al 2003.

\subsection{Outline of this paper}
This is the fourth paper of this series and puts a new emphasis on regions in
which the ambient pressure decreases with height like $z^{-4}$. It also gives
detailed analytical solutions of the $dynamical$
problem for the first time. 
Papers II and III concentrated on why there are collimated jets at all. Here
we concentrate on the dynamic magnetic configurations generated.

Paper I (Lynden-Bell \& Boily 1994) showed that when an
inner disk was rotated by $208$ degrees relative to an outer disk
field lines that had connected them splayed out to infinity in the
absence of any confining external pressure.  At greater angles there
was no torque as the inner and outer disks were magnetically
disconnected.

Paper II (Lynden-Bell 1996) demonstrated that inclusion of a weak
uniform external pressure led to a strong collimation after many turns
with the magnetic field creating towers with jet-like cores whose
height grew with each turn.  Again the inner disk was rotated rigidly
relative to the outer disk.

Paper III (Lynden-Bell 2003) was a refined version of a conference
paper (Lynden-Bell 2001). In these the differential rotation of the
accretion disk and external pressure variation with height were
included and the shape of the magnetic cavity was calculated as a
function of time. However the fields were not calculated in detail and
the treatment used a static external pressure which was assumed to
fall less rapidly than $z^{-4}$.

In this Paper IV we show that these quasi-static models are actually
dynamically correct provided that the motions generated in the field
lines by the twisting of the accretion disk never become relativistic,
but we then explore the consequences of the magnetic cavity expanding
into a region where the pressure variation approaches $z^{-4}$. This
results in a dramatic acceleration of the top of the magnetic cavity
along a cone whose angle gradually narrows at greater distances. The
sudden expansion out to infinity when the twist exceeded a critical
angle, found in paper I when there was no confining medium, is still
present in modified form when beyond some height the external ambient
pressure falls as $z^{-4}$ or faster; once the field has penetrated to
that region there is no longer a quasi-static configuration for the
system to go to, so the jet accelerates to reach either dynamic
ram-pressure balance whenever the background density falls less fast
than $z^{-6}$ or failing that relativistic speeds.

\subsection{Dynamics from statics}
In the standard MHD approximation the displacement current is
neglected so curl${\bf B}=4\pi {\bf j}$.  If the magnetic field
dominates over any other pressure or inertial forces, then, neglecting
those, the magnetic force density has nothing to oppose it, so ${\bf
j\times B=0}$ and we deduce that the currents flow along the lines of
force.  This implies that ${\bf j=}\tilde \alpha{\bf B}$ and as both
div${\bf B}$ and div${\bf j}$ are zero $\tilde \alpha$ is constant
along the lines of force.

We shall be considering problems with the normal component of magnetic
field specified on an accretion disk at $z=0$ and the field does not
penetrate the other boundary where an external pressure $p(z)$
balances $B^2/8\pi $. The past motion of the accretion disk has
produced a twist in the field-lines which emerge from the disk and
return thereto further out. Those with past experience will know that
the above conditions supplemented by expressions for $B_n$ on the
disk, the twists $\Phi$ on each line and $p(z)$, serve to define the
problem, so the magnetic field is then determined. The other Maxwell
equations curl${\bf E}=-\partial{\bf B}/\partial ct$, and that giving
div${\bf E}$, are not needed in the determination of the magnetic
field. Notice that all the equations used to find the magnetic field
do not involve the time. Thus if we specify the twist angles together
with the normal field component on the disk and the boundary pressure
$p(z)$ at any time, the whole field configuration at that time is
determined. Now let us suppose that we know how to solve this static
problem but that $B_n(R),\Phi(R)$ and $p(z)$ are continuously
specified as functions of time. Then the solution for the magnetic
field in this dynamic problem is found by merely taking the sequence
of static problems parameterised by $t$. Such a procedure will give us
${\bf B(r},t)$ but in the dynamic problem the motions of the lines of
magnetic force generate electric fields that must be found
also. However these can be found almost as an afterthought because we
can determine how the lines of force move and their motion is the
$c{\bf E\times B/B^2}$ drift. Assuming perfect conductivity there is
no component of ${\bf E}$ along ${\bf B}$ so ${\bf E=B\times
u}/c$. Our knowledge of ${\bf u}$ on the accretion disk tells us how
the lines move everywhere, which in turn tells us the electric field.

We conclude that the crux of the problem lies not in difficult and
dangerous dynamics but in the staid simplicity and safety of
statics. That said we need whole sequences of static solutions that
allow us to turn up the twists and parameterise the external
pressures. Even the static problem is no walkover and we would have
found it impossible to get general solutions were it not for the
energy principle that the magnetic field adopts the configuration of
minimum energy subject to the flux, twist and pressure constraints
imposed at the boundary.

Here we have already demonstrated why an evolution of the magnetic
field structure through quasi-static models gives the solution to the
dynamic problem. Section 2 details the specification of the relevant
static problem and the methods of solving it. In Section 3 we solve it
developing further the approximate method of paper III. This gives us
the mean fields within magnetic cavities whose shapes we calculate for
any prescribed external pressure distribution. Emphasis is placed on
solutions that access regions with $p$ falling like $z^{-4}$ and the
very fast expansions then generated.  We then generalise our results
to allow for a dynamic ram-pressure and discover the shapes of
inertially confined jets. In Section 4 we calculate the detailed
magnetic fields within the cavity. Section 5 finds the electric fields
generated as the magnetic cavity grows and categorises the types of
solution. Section 6 gives exact solutions of special cases with the
dynamical electric field also calculated.

\section{The Magnetic Problems to be Solved}
A magnetic flux $P(R_i,0)$ rises out of an accretion disk on $z=0$ at
radii up to $R_i$ . The lines of magnetic force on a tube encircling
the flux $P$ eventually return to the disk at an outer radius $R_o(P)$
after a total twist around the axis of $\Phi(P)$. The magnetic field
above the disc is force free with current flowing along the field
lines and there is negligible gas pressure within the
magnetic-field-dominated cavity. However the magnetic cavity is
bounded by a surface at which an external pressure $p(z)$ is
specified. Later we shall consider the case of a dynamical pressure
$p(z,t)$.  Our problem is to find the magnetic field and the shape of
the cavity containing it when the functions $P(R,0),\Phi(P)$ and
$p(z)$ are specified.  Axial symmetry is assumed. We think of $\Phi$
as due to the disk's past differential rotation and sometimes write
$\Phi(P)=[\Omega_d(R_i)-\Omega_d(R_o)]t=\Omega(P)t$ where the suffix
$d$ refers to the rotation of the disk itself. The equation div${\bf
B}=0$ implies that the magnetic field components in cylindrical polar
coordinates $(R,\phi,z)$ may be written in terms of the flux function
$P(R,z)$ which gives the flux through a ring of radius $R$ at height
$z$, and the gradient of the azimuthal coordinate $\phi$, in the
form,
\begin{equation} \label{1}
{\bf B}=(2\pi)^{-1}\nabla P\times\nabla\phi +B_{\phi}\hat\phi.
\end{equation} 
The force-free condition $4\pi{\bf
j\times B}=$curl${\bf B\times B}=0$ then tells us that $B_\phi$ takes
the form,
\begin{equation} \label{2}
B_\phi=(2\pi R)^{-1}\beta(P).
\end{equation}
The function $\beta$ is constant along each field line (so it is a function
of P) and must be determined from the solution so that the total twist on
that field line is $\Phi(P)$. Finally the azimuthal component of the force
free condition yields the equation
\begin{equation} \label{3}
\nabla^2 P-\nabla \ln R^2.\nabla P =R{\partial\over \partial
R}({1\over R}{\partial P\over \partial R})+{\partial^2P\over\partial
z^2}=-\beta ^\prime(P)\beta(P)=-8\pi^2Rj_\phi.
\end{equation}
This is to be solved within an unknown surface S given by $R=R_m(z)$ in
which the field lies and on which ${\bf B^2}=8\pi p(z)$.

\noindent The difficulties of this problem are:

- the equation is non-linear

- the function $\beta$ in it is not known and can only be determined from
$\Phi(P)$ via the solution. 

- the bounding surface S is unknown.

Luckily there is a different way of tackling these magnetostatic
problems. The energy of the magnetic field and of the external
gas-pressure must be a minimum subject to the flux and twist
conditions on the accretion disk. The pressure energy stored in making
a cavity whose area at height $z$ is $A(z)$ against an external
pressure $p(z)$ is
\begin{equation} \label{4}
W_p=\int p(z)A(z)dz. 
\end{equation}
The energy principle that applies even outside axial symmetry is that $W$
must be a minimum, where 
\begin{equation} \label{5}
8\pi W=\int\left [\int\int (B_R^2+B_\phi^2+B_z^2)Rd\phi
dR+8\pi pA\right]dz,
\end{equation}
and ${\bf B}$ satisfies the flux and twist conditions on $z=0$. Two
exact theorems follow; they were proved in paper III by expanding a
horizontal slice through the configuration first vertically and then
horizontally. Earlier more specialised versions appeared in papers I
and II.  These theorems are true even without axial symmetry. Defining
averages $<..  >$ over a horizontal plane at height $z$,
\begin{equation} \label{6}
<B_R^2>+<B_\phi^2>=<B_z^2>+8\pi\left( p(z)+[A(z)]^{-1}\int_z
A(z')[dp/dz']dz'\right);
\end{equation}
the final integral from $z$ up is negative whenever pressure falls at height.
Minimum energy for horizontal displacements of the slice gives,
\begin{equation} \label{7}
<B_z^2>=8\pi p(z)-(4\pi/A)dW_0(z)/dz,
\end{equation}
where
\begin{equation} \label{8}
4\pi W_0=\int\int B_RB_zR^2d\phi dR,
\end{equation}
and the integration is over the plane at height $z$.

We shall show presently that after much twisting the magnetic configuration
becomes very tall as compared to its width. As no extra radial flux is
created the radial field lines become widely spread out over the height of
the resultant tower and the gradients with height likewise become small. When
we neglect terms involving $B_R$ equation (\ref{7}) becomes
\begin{equation} \label{9}
<B_z^2>=8\pi p(z),
\end{equation}
and with a similar neglect in the use of equation (\ref{6}) we find
\begin{equation} \label{10}
<B_\phi^2>/(2-s) = <B_z^2> = 8\pi p(z),
\end{equation}
where we have defined a dimensionless s (positive when pressure falls at
greater height) by,
\begin{equation} \label{11}
 s(z)=-\int_z A(z')(dp/dz')dz'/[A(z)p(z)].
\end{equation}
Notice that $s=0$ for the constant pressure case and that $(1-s)/s$ is the
constant $d\ln(A)/d\ln(p)$ when $A$ the cross sectional area varies as a power
of $p$.

Two possible approaches to solving this problem are:

\noindent I. FORWARDS METHOD
 
We use the energy variational principle choosing such simple trial functions
that the boundary conditions can be applied and the resulting variational
equations can be solved. We can then solve the problem for all choices of the
prescribed functions $P(R,0),\Phi(P)$ and $p(z)$ but the accuracy of the
solution is limited by the imposed form of the trial function.

\bigskip
\noindent II. BACKWARDS METHOD

Here we solve the exact equation (\ref{3}) but make special choices of
$\beta(P)$ and of the surface S so that we can solve the problem. Once
we have the solution we discover the functions $P(R,0),\Phi(P),$ and
$p(z)$ for which we have the solution. While this method is limited to
a very few special cases, at least for them the solutions are
exact. This enables us to check the accuracy and the validity of the
more general solutions given by the Forwards method.

\section{Solution by Variational Principle}
This method was developed in paper III and with an extremely crude but
instructive trial function it was used in paper II. There we showed
that if each field line turned N times around the axis and the
magnetic cavity was taken as a cylinder of height $Z$ and radius $R$
then, if the total uprising poloidal flux was $F$, very crude
estimates of the field components are: $B_z=2F/(\pi
R^2);B_R=F/(\sqrt2\pi RZ);B_\phi =NF/(RZ)$. Squaring these estimates,
adding the external pressure $p$ and multiplying by the volume $\pi
R^2Z$ we find $8\pi^2W=F^2[(1/2
+N^2\pi^2)Z^{-1}+(4R^{-2}+8\pi^3pR^2F^{-2})Z]$ where $p$ is assumed
independent of $z$. Minimising over $R$ gives $R=(2\pi^3p/F^2)^{-1/4}$
so R is determined by the external pressure and the flux. With this
result the final round bracket in $W$ reduces to $8R^{-2}$ and
minimising $W$ over all $Z$ we find
$4Z/R=\sqrt{1+2N^2\pi^2}\rightarrow \sqrt2\pi N$ which shows a
remarkable collimation which improves with every turn!

The method of paper III involved a much improved trial function which allows
each field line, $P$, to attain whatever maximum height, $Z(P)$ it likes,
allows for the different total twists $\Phi(P)$ of the different field lines
and properly accounts for the variation of external pressure with height.
Because of the great heights of the magnetic towers generated after many
turns, most of the field energy is high above the accretion disk and the
detailed distribution of the flux in $P(R,0)$ no longer plays a part in the
distant field. At each height, z, mean fields are defined by
\begin{equation} \label{12}
\overline B_\phi(z)=R_m^{-1}\int_0^{R_m}B_\phi(R,z)dR,
\end{equation}
where $R_m(z)$ is the radius of the magnetic cavity at height $z$, and
$A(z)=\pi R_m^2$. Also
\begin{equation}\label{13}
\overline{ |B_z|}=A^{-1}\int_0^{R_m}|B_z|2\pi
RdR=A^{-1}\int|\partial P/\partial R| dR=2P_m/A.
\end{equation}
The last expression arises because $P$ is zero at both $R=0$ and $R=R_m$ and
achieves its maximum $P_m(z)$ at an intermediate point. We also define
\begin{equation} \label{14}
I^2=<B_z^2>/ \overline{| B_z|}^2,
\end{equation}
and 
\begin{equation} \label{15}
J^2=<B_\phi^2>/\overline B_\phi^2;
\end{equation}
although $I$ and $J$ are in principle functions of height, for tall
towers we expect them to settle down to some typical values which we
determine later. In what follows we neglect any variation of J with
height; variation of I does not change the result. We now explain the
basis of the trial function used.

If a poloidal flux $dP$ is twisted once around the axis, it generates
a toroidal flux $dP$. If its twist is $\Phi(P)$ the toroidal flux
generated is $(2\pi)^{-1}\Phi(P)dP$. First consider distributing this
toroidal flux uniformly over the height $Z(P)$ to which this field
line reaches as was done in paper III. Unlike the use of $Z$ in the
crude calculation in its new meaning it depends on the $P$ of the
field line. In a small height interval $dz$ the element of flux $dP$
contributes a toroidal flux $[2\pi Z(P)]^{-1}\Phi(P)dPdz$ whenever the
height is less than $Z(P)$. The total toroidal flux through $dz$ is
contributed by all lines of force that reach above that height; that
is by those with $P<P_m(z)$, where $P_m(z)$ is the maximum value that
$P$ achieves at height $z$. Hence there is a toroidal flux through the
area $R_m dz$ of

$$R_m\overline B_\phi dz=(2\pi)^{-1}\int_0^{P_m}(\Phi/Z)dP dz;$$
$\Phi(P)$ is specified in terms of the accretion disk's twist. $Z(P)$
is to be varied. However while this makes a possible trial function it
is not generally true that the flux is so distributed. Indeed in the
limiting case of the exact self-similar solutions discussed in the
next paper the twist is strongly concentrated toward the top of each
field line. The extreme alternative is found by putting all the twist
at the top of each field line.  Then the toroidal flux in any height
increment $dz$ depends on the twist of those field lines whose tops
lie in $dz$ so $R_m\overline B_\phi
dz=(2\pi)^{-1}\Phi(P_m)(-dP_m/dz)dz$. However the truth must lie
between the extremes of uniform twist and all twist at the top of each
line. We therefore take the geometric mean of these two expressions
for the toroidal flux distribution and obtain:
\begin{equation}  \label{16}
R_m\overline{B_\phi}=(2\pi)^{-1}[\Phi(P_m)(-dP_m/dz)\int_0^{P_m}
(\Phi/Z) dP]^{1/2}.
\end{equation}
Now $Z(P)$ is in the variational principle and $P_m(z)$ has the
property that $P_m[Z(P)]=P$, so $P_m$ is the functional inverse of
$Z(P)$; so whenever $Z(P)$ is varied $P_m(z)$ automatically varies in
concert.  We now omit the $B_R^2$ term in the energy principle as it
is much smaller than the others once the towers grow tall, cf the
crude estimate above equation (\ref{12}). Writing $W$ in terms of the
expressions discussed above
\begin{equation} \label{17}
8\pi W=\int \left[(4\pi)^{-1} J^2 \Phi(P_m)
(-dP_m/dz)\int_0^{P_m(z)}\Phi/Z.dP +4I^2P_m^2/A+8\pi p(z)A\right]dz
\end{equation}
Varying $A(z)$ gives 
\begin{equation} \label{18}
A=\pi R_m^2=IP_m/\sqrt{2\pi p(z)}
\end{equation}
which using (\ref{13}) and (\ref{14}) is equivalent to $<B_z^2>=8\pi
p$, as found in equation (\ref{9}). Putting this expression back into
equation(17) we find that the last two terms there combine to make
$P_md\Pi/dz$ where $\Pi=\int_0^z4I\sqrt{2\pi p(z')}dz'$. We now
integrate both terms in $W$ by parts and then rename the dummy
variables in the integrals that result from the first term; these
operations give
\begin{equation}8\pi \label{19}
W=(4\pi)^{-1}J^2\int_0^F[\Phi(P_m)/Z(P_m)]\int_{P_m}^F\Phi
(P)dPdP_m+\int_0^F\Pi dP_m,
\end{equation}
where $\Pi$ is merely $\Pi(z)$ re-expressed as a function of $P_m$
\begin{equation} \label{20}
\Pi(P_m)=\int_0^{Z(P_m)}4I\sqrt{8\pi p(z)}dz
\end{equation}
Most remarkably our adopted geometric mean between the uniform twist
and top- twist cases has resulted in a $W$ of precisely the form
obtained in the uniform twist case of paper III except that the first
term is now exactly half of its former value. Since this clearly
reduces $W$ the new trial function is clearly better than the old in
that it brings us closer to the true minimum. Minimising $W$ over all
choices of $Z(P_m)$ we find from (\ref{19}) and (\ref{20})
\begin{equation} \label{21}
Z^2\sqrt{8\pi p(Z)}=[J^2/(16\pi I)]P_m\overline \Phi(P_m)^2
\end{equation}
where
\begin{equation} \label{22}
\overline\Phi^2=\Phi(P_m)P_m^{-1}\int_{P_m}^F\Phi(P)dP
\end{equation}
Equation (\ref{21}) gives the function $Z(P_m)$ and its inverse $P_m(z)$ in
terms of the given $p(z)$ and $\Phi(P)$. In line with section two's
introduction we write $\Phi=\Omega t$ and $\overline\Phi=\overline{
\Omega} t$.  From equation (21) we notice that as a result of this $t$
dependence $\overline\Phi$ and $Z$ grow with $t$ at each value of
$P_m$. This growth is at constant velocity if the external pressure is
independent of height.  However if that pressure falls off with height
the velocity $dZ/dt$ accelerates. e.g. if $p\propto (a+Z)^{-n}$ with
$n=2$ then $Z$ has nearly constant velocity until it reaches '$a$' but
at greater heights it behaves as $t^2$ with constant acceleration. The
shape of the magnetic cavity at each moment is given by plotting $Z$
against $R_m(Z)$. We find the relationship by substituting $P_m$ from
equation (18) into equation (21) to give the equation for $Z$ as a
function of $R_m$. Viz $Z=[J/(4\sqrt{2}I)]R_m\overline\Omega [\pi
I^{-1}R_m^2\sqrt{2\pi p(Z)}].t, $ where $\overline\Omega $ is
evaluated at the value of $P_m$ indicated in its square
bracket. Dividing equation (21) by equation (18) we find the
collimation at any given $P_m$ is just
\begin{equation} \label{23}
Z/R_m=[J/(4\sqrt{2}I)].\overline\Omega(P_m)t,
\end{equation}
which still grows linearly with time even when the external pressure
decreases with height. However $p(z)$ should not decrease too fast
else equation (\ref{21}) will not have a sensible solution. To
understand this we see from equation (\ref{22}) that
$P_m\overline\Phi^2$ decreases as $P_m$ increases since the twist
$\Phi(P_m)$ is greatest for the field lines rising nearest the centre
of the disk. Also $P_m$ decreases at greater heights since less
magnetic flux reaches there. Hence $P_m\overline\Phi^2$ increases with
height $Z$. 
\begin{figure}
\begin{center}
\includegraphics[scale=0.6,angle=0.0,clip]{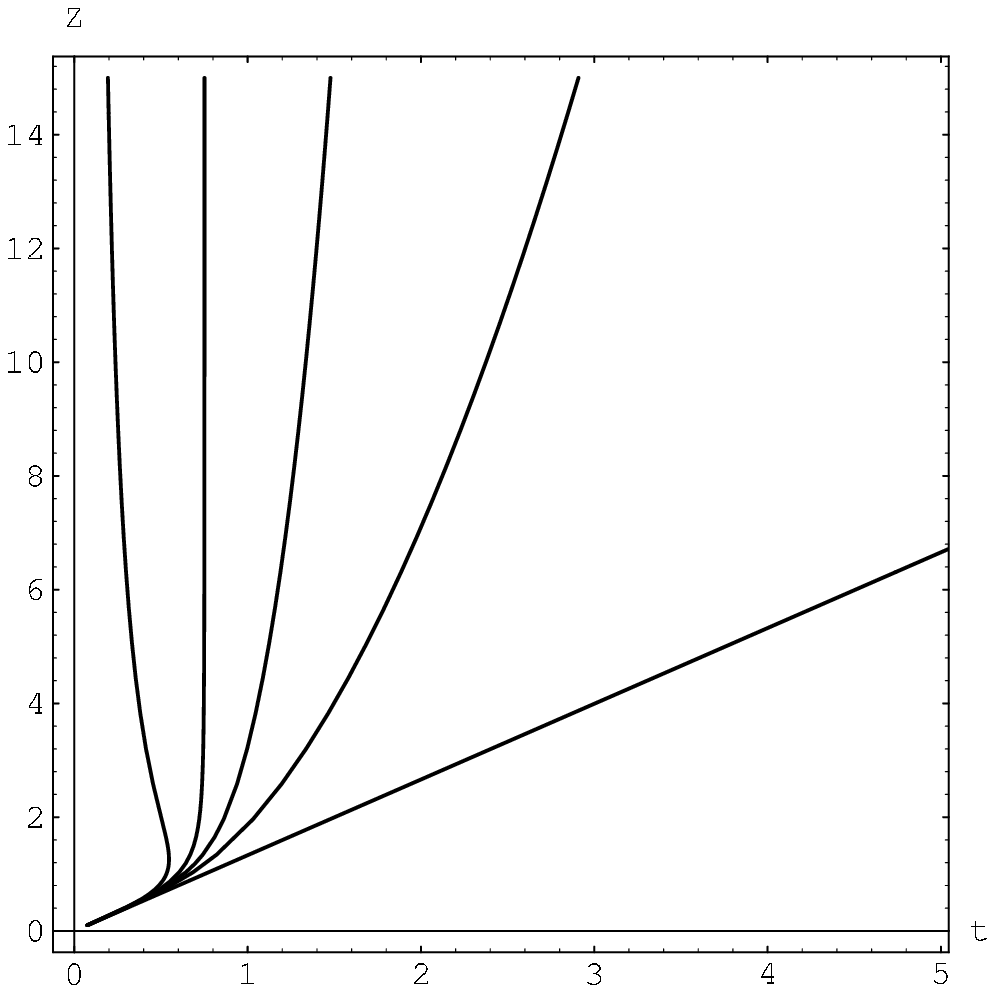}
\includegraphics[scale=0.6,angle=0.0,clip]{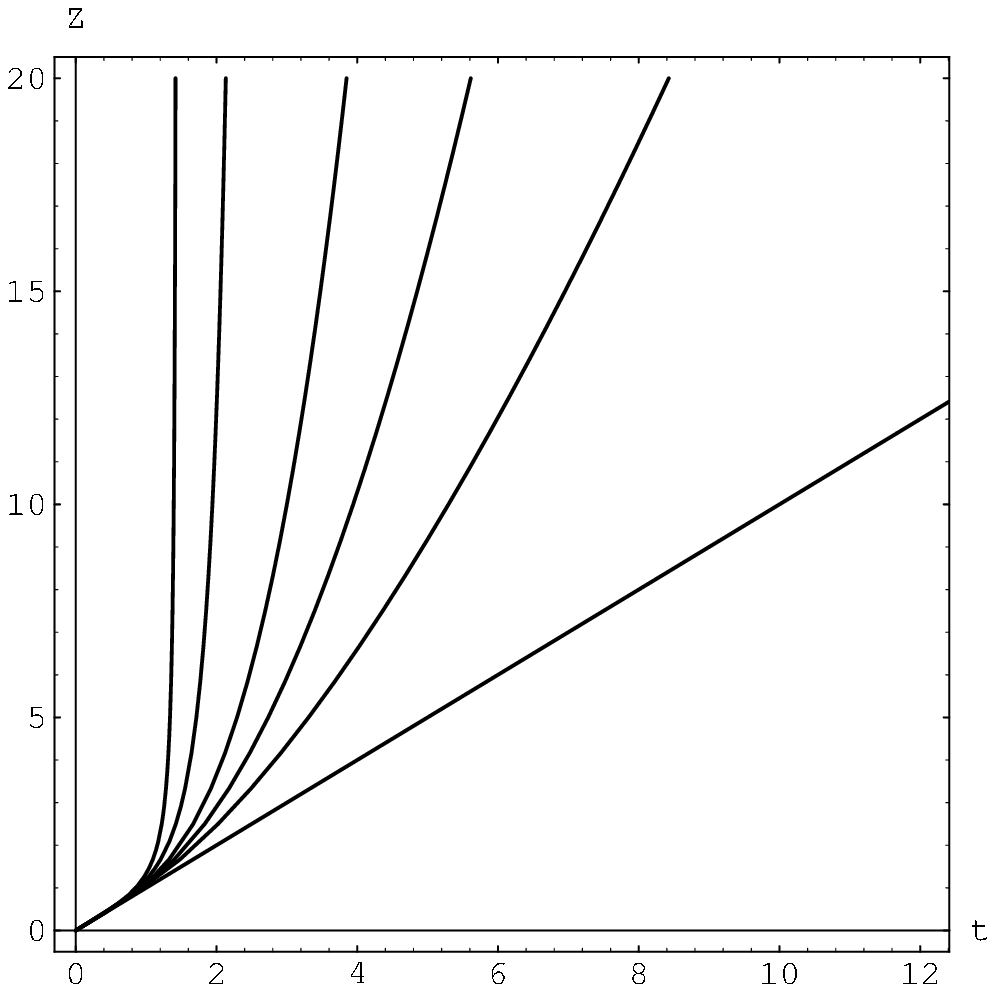}
\caption{Figure 1 shows the time evolution of the height of the jet in
different pressure environments. On the left for the lowest graph the
jet penetrates a constant pressure environment, n=0, at constant
speed; the higher graphs are for n=2,3,4,6, and n=2 has almost
constant velocity up to Z=a but feels the pressure decrease and
thereafter proceeds at almost constant acceleration. At higher n the
acceleration increases and there are no continuing solutions with
magnetic pressure in balance beyond n=4. In fact the n=6 curve turns
back unphysically to earlier times because the external pressure is
too weak to withstand the magnetic field at larger heights. In reality
the magnetic field springs outward at such a speed that dynamic
ram-pressure comes into play. On the right we see the equivalent
figure for ram-pressure with density profiles m=0,2,3,4,6,8 see
equation (\ref{26}) }
\end{center}
\label{fig1}
\end{figure}
This must be true of the left hand side of equation
(\ref{21}) too and indeed it is obviously so when $p$ is constant so
there is then a sensible solution. However should $p(Z)$ decrease
faster than $Z^{-4}$ the left hand side of (\ref{21}) would decrease
with $Z$ so there would be no such solution. If the cavity accesses
regions in which $p$ decreases a little less fast than $z^{-4}$ then
the field rapidly expands outwards and this gives a most interesting
jet model. For example if $p\propto(a+z)^{-(4-\delta)}$ then $Z$ would
grow with a very high power of $t$, $t^{2/\delta}$ at each $P_m$ so
high expansion speeds would be achieved quite soon; however ram
pressure will increase the effective pressure at the jet's head. This
is explored in a later section. When relativistic speeds are achieved
our approximation fails but analytical progress with relativistic jets
can now be achieved following the lead of Prendergast (2005). For
static pressures with $p=p_0/[1+(z/a)]^n$ we illustrate the motion of
the jet-head in Figure 1.
For $Z<a$ the velocity is almost constant
but as the jet encounters the decrease in pressure it accelerates. For
$n=2$ the acceleration becomes uniform, but for larger $n$ it
increases and for $n=4$ infinite speeds would occur in finite time if
our equations still held. For $n>4$ the graph turns back so the
solutions indicate an infinite speed and then turn back giving no
solutions for later times. These results are of course modified when
ram-pressure is included as described later.

\subsection{CAVITY SHAPES FOR SPECIFIC MODELS}

We have the solution for any specified distributions of $p(z)$ and for
any twist $\Phi(P)=\Omega(P)t$.  However before we can draw any cavity
shapes we must also specify $\Omega(P)$ or equivalently
$\overline{\Omega}(P)$. These two are connected through the definition
$P[\overline\Omega(P)]^2=\Omega(P)\int_P^F\Omega(P')dP'=-(1/2)d/dP[\int_P^F\Omega(P')dP']^2$.
This relationship is easily inverted. Evidently
$\int_P^F\Omega(P')dP'=[\int_P^F2P\overline{\Omega}^2dP]^{1/2}$ so
finally
$\Omega(P)=P\overline{\Omega}^2[\int_P^F2P\overline{\Omega}^2dP]^{-1/2}.$
We expect the behaviour of $\Omega(P)$ near $P=F$ should be
proportional to $(F-P)^2$ since on the disk $B_z=0$ there. From the
definition above this implies that $\overline{\Omega}^2$ should be
proportional to $(F-P)^3$ near there. It is also true that if $\Omega$
tends to $\Omega_0$ as $P$ tends to zero then $\overline{\Omega}$ will
be proportional to $P^{-1/2}$ near there.

\begin{figure}
\begin{center}
\includegraphics[scale=0.39,angle=0.0,clip]{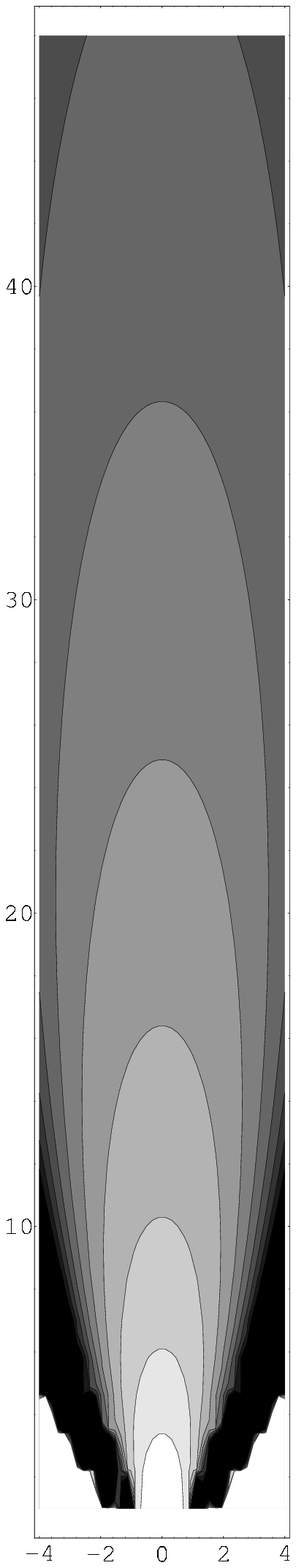}
\includegraphics[scale=0.4,angle=0.0,clip]{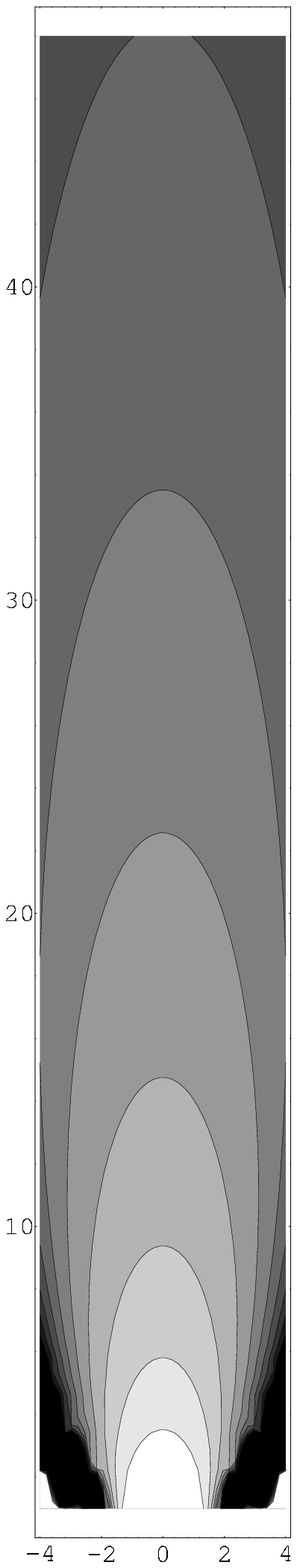}
\includegraphics[scale=0.44,angle=0.0,clip]{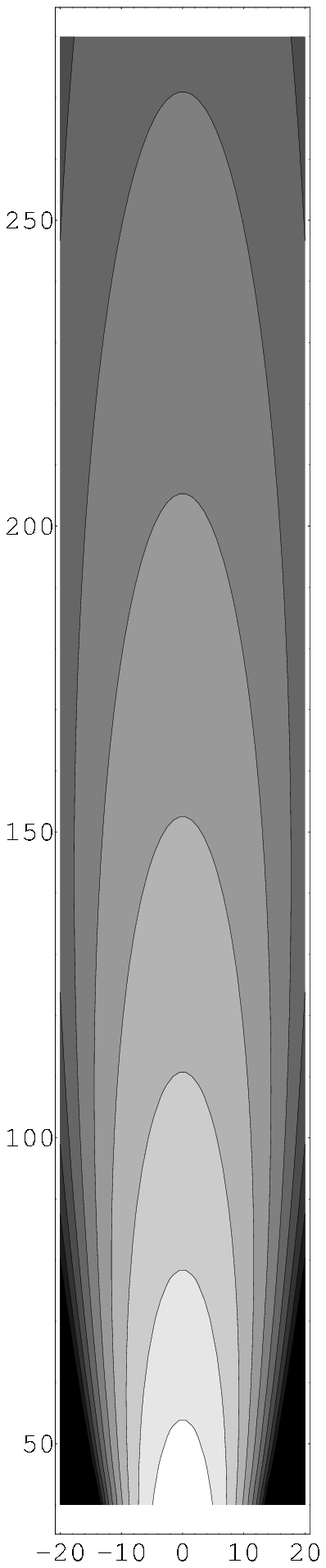}
\includegraphics[scale=0.4,angle=0.0,clip]{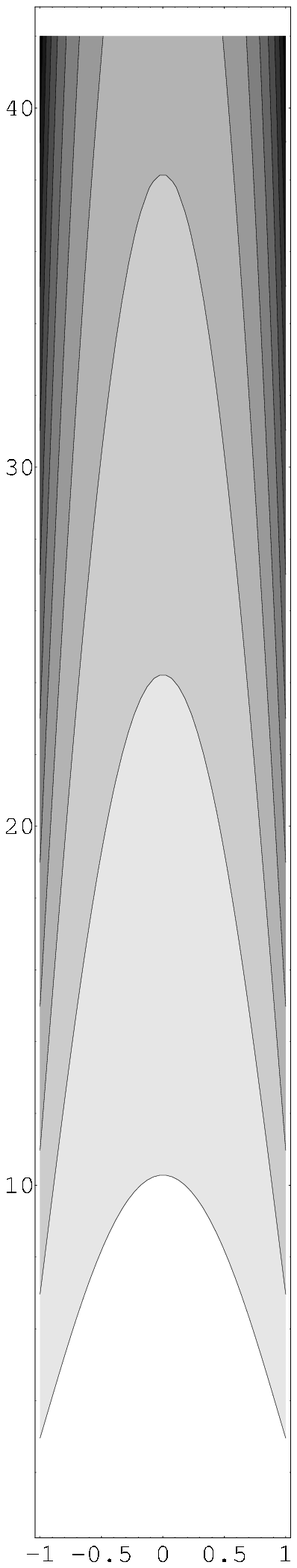}
\caption{The two figures on the left give the temporal evolution of the cavity
shapes for the Dipole (left) and Simple models for the n=3 pressure
distribution. The times corresponding to a given jet height are given in
figure 1. Next we have the Simple model at a later time (notice the scale
change). These are contrasted with the evolution of a magnetic cavity in a
constant pressure environment n=0, on the right; here at each time the cross
sectional area is always smaller at greater height.}
\end{center}
\label{fig2}
\end{figure}
The simplest expression with these properties is the Simple Model $
\overline{\Omega}=(2)^{-1/2}\Omega_0(P/F)^{-1/2}(1-P/F)^{3/2}$. This
steadily falls from its central value and corresponds to
$\Omega(P)=\Omega_0(1-P/F)$.

A model with a more physical motivation is a uniformly magnetised
rapidly rotating star or black hole giving an initial flux
distribution on the disk $P(R,0)=FR^2/a^2$ for $R<a$ and $P=Fa/R$ for
$R>a$. This Dipole Model has a sudden reversal of field at $R=a$. We
combine this with a rotation
$\Omega_d=\Omega_*---R<a;\Omega_d=\Omega_*(R/a)^{-3/2}---R>a$.
Setting $Y=P/F, \Omega=\Omega_*[1-Y^{3/2}]$ and on integration
$\overline\Omega(P)=\Omega_*Y^{-1/2}\sqrt{(1-Y^{3/2})[3/5-Y+(2/5)Y^{5/2}]}$.

The shapes of some of the dynamic magnetic cavities generated
by combining these $\Omega(P)$ distributions with simple ambient external
pressure distributions $p(z)$ are illustrated in Figure 2.

Figure 2 shows the time evolution of the magnetic cavity for the $n=3$
pressure distribution for the Dipole and the Simple models. They do
not differ much. Very different cavities and velocities arise when the
pressure distribution in the ambient medium is changed. In particular
larger $n$ albeit less than 4 gives a fatter jet at a given length and
a longer jet at a given time as illustrated in figure 2, however the
collimation as determined by the length to width ratio is governed by
equation (\ref{23}) and so remains about the same at a given time.
With the shapes of the cavities known it is now possible to calculate
their areas at each height for each external pressure and thus discover
how $s(z)$ varies with height. This is interesting as, from
(\ref{10}), $<B_\phi^2>/<B_z^2>=2-s$. For $z>>a$ the distribution of
$2-s$ with $(z/Z_h)^2$ is given for various values of $n$ using the
Simple Model.  These graphs illustrate that the twist is less for the
higher values of $n$ but for them it is more concentrated toward the
top of the jets. The integrations for equation (\ref{11}) were
performed by expressing the area as a function of pressure via
equations (\ref{18}), (\ref{21}) and (\ref{23}). The final function
for $z>>a$ which is plotted is
$2-s=(1/2)[(4-n)/(n-1)]x^b[(1-x^c)/(1-x^b)]$ where $x^2=(z/Z_h)
;b=(4-n)/3;c=4(n-1)/3 $
\begin{figure}
\begin{center}
\includegraphics[scale=1.0,angle=0.0,clip]{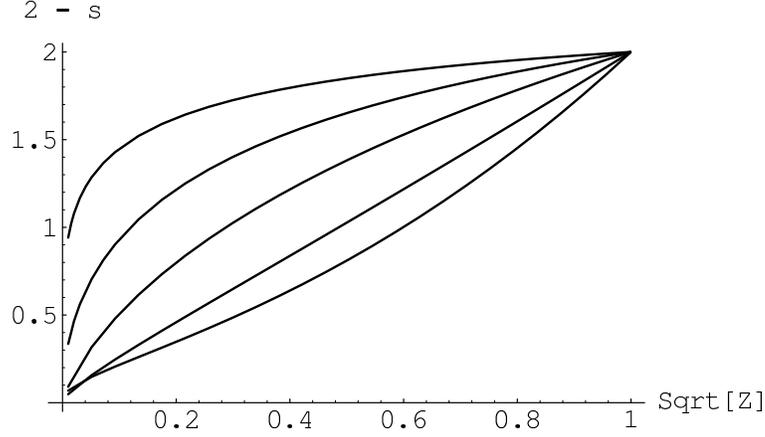}
\caption{$<B_\phi^2>/<B_z^2>$ plotted as a function of $\sqrt{z/Z_h}$ for
pressure-confined Simple-model jets  with from the top n=0.2,0.5,1,2,3. Both
field components become zero at the jet's head}
\end{center}
\label{fig3}
\vspace{-0.2cm}
\end{figure}

\subsection{FAST JETS -- RAM PRESSURE}
Our most suggestive finding thus far is that there are NO quasi-static
solutions of large total twist $ \Phi$ when the external pressure
falls like $z^{-4}$ or faster. This result is easily understood; A
purely radial field in a bottom-truncated cone $r > a$ of solid angle
$\omega$ will fall like $r^{-2}$. If the total poloidal flux both
outwards and inwards is $F$ the magnetic field would be $2F/[\omega
r^2]$ and it would deliver a pressure $F^2/[2\pi \omega^2 r^4]$ on the
walls. The total field energy would be $F^2/[2\pi\omega a]$ and an
equal amount of work would be needed to inflate the magnetic cavity
against the external pressure so the total energy would be $F^2/[\pi
\omega a]$. If instead we had a pure potential field with the same
flux its field would fall as $r^{-(l+2)}$ so its energy would be about
$F^2/[2\pi(2l+1)\omega a]$ where $l$ is the order of the Legendre
polynomial that fits into the solid angle; so the energy of the purely
radial field to infinity and back is only about $2l+1$ times the
energy of the potential field. In practice $l=(4\pi /\omega)^{1/2}.$
In paper V we find that a total twist of the upgoing flux relative to
the downcoming flux, of $\pi/sin[\theta_m/2]$ is sufficient for the
flux to extend to infinity within a cone of semi-angle $\theta_m$.
When the pressure falls with a power a little less negative than minus
four we already demonstrated that continued twisting leads to the top
of the tower or jet head advancing with a very high power of the
time. However that was on the basis of a static pressure which will be
enhanced by the dynamic ram-pressure once speeds comparable with the
sound speed are achieved.  Applying Bernoulli's equation in the frame
with the jet-head at rest we have $ v^2/2+[\gamma/(\gamma-1)]p/\rho =
$constant, so at the stagnation point $p_s = p[1+(\gamma-1)/2\gamma
.\rho v^2/p]^{\gamma/(\gamma-1)}$. If $p >>\rho v^2$ we get
$p_s=p+\rho v^2/2$. Although that is much used we are more interested
in the general case. When the flow becomes supersonic a stand-off
shock develops so the above formula needs to be modified. Just behind
the bow shock the pressure and density are given in terms of the
upstream Mach number $M$ by $p_2= p(\gamma+1)^{-1}[2\gamma
M^2-(\gamma-1)]; \rho_2=\rho (\gamma+1)M^2/[(\gamma-1)M^2+2]$; also $
\rho_2v_2=\rho v.$ Applying Bernoulli's theorem after the shock
results in a formula which for Mach numbers of two or more is well
approximated by $p_s=\Gamma(\gamma)\rho v^2$. Here
$\Gamma=[(\gamma+1)/2]^{(\gamma+1)/(\gamma-1)}\gamma^{-\gamma/(\gamma-1)}$
which is $1,0.93,0.88$ for $\gamma=1,4/3,5/3,$ respectively. The full
formula for $p_s$ has a further factor on the right
$[1-(\gamma-1)/(2\gamma M^2)]^{-1/(\gamma-1)}$ which clearly tends to
one at large $M^2$. For $\gamma=4/3$ it ranges between $1.49 --> 1.05$
as $M$ ranges from one to three. For $\gamma=5/3$ the range is from
$1.41 --> 1.03$.  A useful global formula that somewhat approximates
the trans-sonic pressure but is good in both the static and strongly
supersonic limits is
\begin{equation} \label{24}
p_s=p+\rho v^2.
\end{equation}
This is the stagnation pressure that will be felt at the head of the jet.
Away from the head the pressure is reduced both because the magnetic cavity
expands less rapidly in the direction of its normal there and because a
considerable fraction of the velocity is now parallel to the surface of the
cavity. We return to this in the paragraph on Inertially Confined Jets.

\subsection{MOTION OF THE JET-HEAD}
From equations (\ref{21}) \& (\ref{22}) the position of the jet head
at time $T$ is $Z_h$ where
\begin{equation} \label{25}
Z_h^2\sqrt{8\pi p_s}=J^2[16\pi
I]^{-1}\Omega(0)\int_0^F\Omega(P)dP.t^2
\end{equation}
where, because we are dealing with the head, we have replaced the
pressure with the stagnation pressure and set $P_m=0$. With the
stagnation pressure given by (\ref{24}) but $\dot{Z}_h$ replacing $v$,
equation (\ref{25}) is now an equation of motion for the jet-head. As
only $Z_h$ occurs in this subsection we shall usually drop the suffix
$h$ which will be understood. When formulae are to be used in other
sections we shall resurrect the suffix so that they can be lifted
unchanged. Squaring (\ref{21}) $p(Z)+\rho(Z)\dot{Z}^2=L^2(t/Z)^4$
where $p(z)$ and $\rho(z)$ are the undisturbed pressure and density at
height $z$ and $L$ is the constant
$(1/2)(8\pi)^{-3/2}J^2I^{-1}\Omega(0)\int_0^F\Omega(P)dP$.  We are
interested in this equation when pressure falls with height. If the
initial velocity given by neglecting the $\dot{Z}^2$ term is subsonic
then initially we shall have results very similar to the quasi-static
case illustrated on the left of figure 1 but all those solutions
accelerate as the pressure decreases so the velocity will become sonic
and then supersonic so that the $\rho(Z)\dot{Z}^2$ term will dominate
over the falling pressure. Now neglecting the pressure and taking the
square root we have $Z^2[\rho(Z)]^{1/2}\dot{Z}=Lt^2.$ Setting
\begin{equation} \label{26}
\rho(Z)=\rho_0[1+(Z/a)^3]^{-m/3},
\end{equation}
so that the density behaves as $z^{-m}$ at large heights, we integrate
to find 
\begin{equation} \label{27}
[1+(Z/a)^3]^{1-m/6}-1=(1-m/6)L_1t^3.
\end{equation} 
where
$L_1=L\rho_0^{-1/2}a^{-3}$.  For all $m$ the solutions for small
heights are $Z=a (L_1)^{1/3}t$ and for large heights
\begin{equation} \label{28}
Z_h=a[(1-m/6)L_1]^{2/(6-m)}t^{6/(6-m)}.......m<6.
\end{equation}
Once again such solutions accelerate but less rapidly; however if the
density ever falls as fast as height to the minus six even the
ram-pressure is too weak and the solutions rise formally to infinite
speed before failing altogether when
$t=[6/(m-6)]^{1/3}F^{1/6}L_1^{-1/3}$. The trouble arises because the
rapid fall in density gives too small a ram-pressure to resist the
acceleration of the jet. In practice either the excess inertia due to
relativistic motion or the fact that the density does not fall below
intergalactic values will avoid this behaviour. In the former case we
need to develop the relativistic MHD jet theory. In the latter we may
use our $m<6$ model modified by replacing the density by the
intergalactic one whenever our formula yields a smaller value, i.e. at
$Z_h>a[(\rho_0/\rho_i)^{3/m}-1]^{1/3}$.  As it reaches this region the
velocity reaches $\dot{Z}_h=[L( \rho_i)^{-1/2}]^{1/3}$ and thereafter
it remains at that value. So $Z_h(t)$ is given by formula (\ref{28})
displayed in Figure 1 (right) until it reaches that region but then
maintains its constant speed.

\subsection{INERTIALLY CONFINED SUPERSONIC JETS}
Landau and Lifshitz in their Fluid Mechanics book paragraph 115 give
an elegant theory of supersonic flow past a pointed body, and the
pressure on the body may be found by using the stress tensor $\rho
(\partial\phi/\partial R)^2$ of the velocity potential given in their
formula 115.3. However there are two drawbacks. Firstly it is not
likely that our jet will constitute a pointed body rather than a blunt
one, even if it could be treated as a body at all, and secondly the
resulting formulae involve integrals over the shape of the body which
we can only discover AFTER the pressure is known.  We shall circumvent
such difficulties while maintaining momentum balance by treating the
medium into which the jet penetrates as ionised dust each particle of
which collides with the magnetic cavity. Taking axes that move with
the top of the field lines labelled by $P$ i.e. with velocity $\dot
Z(P)$ we find a ram-pressure on the cavity wall at height $Z$ of $
2\rho\dot Z^2 cos^2\theta$ where $cos^2\theta=\frac{1}{[1+(\partial
Z/\partial R)^2_t]}$. The factor two arises from the assumption of
specular reflection from the cavity wall and at normal incidence the
formula gives a factor two more than the stagnation pressure at the
jet head. This suggests that a dead-cat-bounce off the cavity wall may
be a better approximation than a specular reflection so we shall omit
the factor two in what follows. However even the resulting formula is
hard to apply in practice so we now proceed to simplify it
further. Most of our jets are not far from parabolic at the front, in
which case $Z_h-Z=\kappa(t)R_m^2$. Then $(\partial Z/\partial
R_m)^2=4(Z_h-Z)^2/R_m^2$. Hence we shall adopt for the dynamic
pressure at $Z(P)$
\begin{equation} \label{29}
p_d=\rho \dot Z(P)^2/[1+4(Z_h-Z)^2/R_m^2]
\end{equation}
To find the shape of a dynamically confined jet we need to solve
equation (\ref{21}) with the dynamic pressure $p_d$ replacing $p$ and
use equations (\ref{23}) and (\ref{26}) for $R_m$ and $\rho(z)$. For
$Z>>a$ equation (\ref{21}) becomes
\begin{equation} \label{30}Z^{2-m/2}\dot Z
\sqrt{8\pi \rho_0a^m/[1+4(Z_h-Z)^2/R_m^2]}=[J^2/(16\pi I)]
P_m\bar\Omega^2t^2.
\end{equation}
Except near the jet-head $Z_h-Z>>R_m/2$ so we may neglect the one and
,using (23) for $R_m$ the above equation simplifies to
\begin{equation} \label{31}
Z^{3-m/2}\dot Z=K_1\bar\Omega^3t^3(Z_h-Z). 
\end{equation}
For an initial orientation we take $Z_h>>Z$. We may then integrate
directly using our former result that $Z_h\propto t^{1/(1-m/6)}$ and
obtain $Z\propto t^{(5-2m/3)/[(1-m/6)(4-m/2)]}$, whence it follows
that $Z$ grows with a higher power of $t$ than $Z_h$ (at least for $Z$
small) indeed $X=Z/Z_h\propto t^{1/(4-m/2)}$. When $Z$ is sizeable the
above overestimates $\dot Z$ so it overestimates $dlnX/dlnt$ which
must in reality lie between zero and $(4-m/2)^{-1}$ which is itself
less than $(1/4)(1-m/6)^{-1}$. We get a better estimate of the
behaviour of $X$ by writing equation (\ref{31}) in the form
\begin{equation} \label{32}
X^{4-m/2}[(1-m/6)(1-X)]^{-1}[1+(1-m/6)dlnX/dlnt]=K_2t,
\end{equation}
where at given $P,K_2$ is constant. Now the second square bracket
above only varies between one and 5/4. If at lowest order we neglect
its variation, we may solve for $t(X)$. We may then evaluate
\begin{equation}  \label{33}
dlnX/dlnt=(1-X)/[4(1-m/8)(1-X)+X],
\end{equation} 
(\ref{32}) then determines $t$ as a function of $X$. Evidently the
$t^{1/(4-m/2)}$ behaviour of $X$ persists approximately until $X$ is
near unity; however $Z_h-Z\propto t^{1/(1-m/6)}(1-X)$ and this
actually grows because with $X$ near unity the second square bracket
in equation (\ref{32}) is one and so $Z_h-Z\propto
[K_2(1-m/6)]^{-1/(1-m/6)}X^{(4-m/2)/(1-m/6)}(1-X)^{-m/(6-m)}$, so as
$X$ grows $Z_h-Z$, the distance from the head actually increases. $Z$
never catches up with $Z_h$ despite the fact that $X$ tends to one.
All the above rests on the premise that $Z_h-Z>>R_m/2$ so we now
investigate the behaviour when $Z$ is close to $Z_h$ so that $X$ is
close to one but not so close that $(1-X)Z/R_m$ has to be small,
Writing equation (\ref{30}) in terms of $X$, but omitting the
$dlnX/dlnt$ term as this vanishes with $X$ near unity, and dividing it
by the same equation for $Z_h$ we find
$1+4(Z_h-Z)^2/R_m^2=X^{6-m}/[K_4(P_m)]^2$ where $K_4$ is $
P_m\bar\Omega^2(P_m)$ divided by its non-zero value when $P_m$ tends
to zero.  At constant $P_m$, $(Z_h-Z)/R_m$ clearly increases as $X$
increases but it tends to the limiting value $(1/2)\sqrt{K_4^{-2}-1}$
as $X$ tends to one. As $K_4$ depends only on $P_m$ the shape of the
jet-head is determined by the $\bar\Omega(P)$ function though its size
and position depend on $t$ also. Thus the whole head of the jet and
all parts with $X$ near one grow self similarly. The shape of the
these parts is even independent of the details of the density fall-off
embodied in $m$ but the size of these parts of the jet is proportional
to $t^{m/(6-m)}$ so the rate of self-similar growth depends on $m$. As
explained above the whole length of the jet grows with a different
power of the time so it is just the part with $X$ close to one that
grows self-similarly. To find the shape of the whole cavity we notice
that elimination of Omega-bar between equations (\ref{23}) and
(\ref{30}) or (\ref{21}) coupled with use of (\ref{33}) leads to an
expression for $P_m$ in terms of $Z,Z_h,R_m,t$
\begin{equation} \label{34}
P_m=
\frac{\pi}{2I\left(1-\frac{m}{6}\right)}(R^2_mZ/t)
\left[1+\frac{(1-m/6)(1-Z/Z_h)}{[4(1-m/8)(1-Z/Z_h)+Z/Z_h]}\right]
\sqrt{{8\pi \rho_0 a^m(Z^3+a^3)^{-m/3}\over[1+4(Z_h-Z)^2/R_m^2]}}
\end{equation}
This value of $P_m$ is substituted into the Omega-bar of equation (23)
to yield the equation relating $Z$ and $R_m$ at each time, $Z_h(t)$
being already known. Thus we get the shapes of the inertially confined
jet cavities.  Figure 4 displays one of these at two different times.
\begin{figure}
\begin{center}
\includegraphics[scale=0.6,angle=0.0,clip]{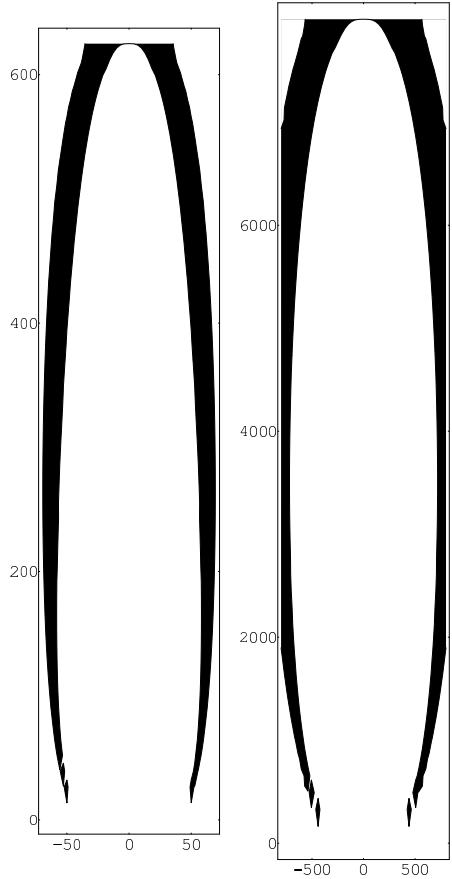}
\caption{An inertially confined jet cavity at two different times, the
inner edge of the black gives the cavity in an m=3 density
distribution for the Simple model magnetic flux. The times to these
heights are given by equation (\ref{27})}
\label{horseshoes}
\end{center}
\label{fig4}
\vspace{-0.2cm}
\end{figure}

\section{THE FIELD IN THE MAGNETIC CAVITY}
Our solution of the variational principle has improved on paper III and
given us dynamical solutions for the cavity's shape and the mean field at
each height. We now seek the detailed field structure within the cavity.
At each height $z$ we define $\lambda$ to be $[R/R_m(z)]^2$ so $P$ may then be
written
\begin{equation} \label{35}
P(R,z)=P_m(z)f(\lambda).
\end{equation}
Since by its definition $P_m(z)$ is the maximum value that $P$ takes
at height $z$, it follows that $f$ achieves its maximum of unity at
each height.  Furthermore, as $P$ is zero both on axis and at the
surface of the magnetic cavity, it follows that $f(0)=0=f(1)$. Thus
$f$, which is positive, is highly circumscribed rising from zero to
one and falling again to zero at one.  Although $f$ may in principle
depend on height (especially near the disk or at the top),
nevertheless so circumscribed a function is unlikely to have a strong
height dependence. We shall make a second approximation that an
average profile will do well enough over at least a local region of
the tower's height. Thus we adopt the form given in equation
(\ref{35}) with $f$ a function of lambda but not of height. Now in the
tall tower limit both $R_m(z)$ and $P_m(z)$ are only weakly dependent
on $z$, so squares of their first derivatives and their second
derivatives may be neglected. Equation (\ref{3}) then takes the form
\begin{equation} \label{36} 
4P_mR_m^{-2}\lambda.
\partial^2f/\partial\lambda^2=-\beta.\partial\beta/ \partial P=
-(\partial\beta^2/\partial\lambda)/(2P_mf^{'}).
\end{equation}
Multiplying by $2P_mf'$ and integrating $d\lambda$ we find
$\beta^2=4P_m^2R_m^{-2}(\int_0^{\lambda}f^{'2}d\lambda-\lambda
f^{'2})$, so $\beta (P)$ is of the form
$\beta[P_mf(\lambda)]=(2P_m/R_m)G(\lambda)$.  Evidently $\beta$ is a
product of a function of $z$ and a function of $\lambda$ but it is
also a function of such a product. It is readily seen that a power law
form for $\beta$ achieves this and we readily prove that the power law
is the only possibility that allows it. Hence we may write
\begin{equation} \label{37}
\beta=C_1P^\nu .
\end{equation}
Thus $\beta'\beta\propto \nu P^{2\nu-1}$. Inserting this into equation
(\ref{36}) and calling the value of $\lambda$ where $f$ achieves its
maximum of one, $\lambda_1$, we find on separating the variables
$\lambda$ and $z$ both
\begin{equation} \label{38}
\lambda d^2f/d\lambda^2=-C^2 \nu f^{2 \nu-1},
\end{equation}
and
\begin{equation} \label{39}
R_m=C_2P_m^{1-\nu},
\end{equation}
where 
\begin{equation} \label{40}
C^2 = C_1^2C_2^2/4.
\end{equation}
Now $P_m(z)$ decreases because not all flux reaches up to great
heights, so equation (\ref{39}) tells us that the radius of the
magnetic cavity decreases there when $\nu<1$ and increases when
$\nu>1$. Of course this only holds once the tower is tall so that its
lateral confinement is due to the ambient pressure rather than the
flux profile on the disk itself.  The constant $C$ is determined by
the requirement that $f=1$ at its maximum because $f(\lambda)$ already
has to obey the boundary conditions at 0 and 1. Sometimes we find it
convenient to use $\alpha=2\nu-1$ in place of $\nu$. We already know
how to calculate the shape of the magnetic cavity.  Knowing $A(z)$ and
$p(z)$ we can calculate $s=-(Ap)^{-1}\int_zA(z')(dp/dz')dz'$ at every
height $z$. We now show that associated with each value of $s$ there
is a profile $f(\lambda)$. Equation (\ref{38}) governs the possible
profiles.  Multiplying it by $-f/\lambda$ and integrating by parts we
find $\int_0^1(f')^2d\lambda
=C^2\nu\int_0^1f^{2\nu}\lambda^{-1}d\lambda$.  However $P_mf'=A(2\pi
R)^{-1}\partial P/\partial R=AB_z$ so the first integral is related to
$<B_z^2>$ and the right hand one is similarly related to
$<B_\phi^2>$. Multiplying both sides by $P_m^2/A^2$ we find
\begin{equation} \label{41}
<B_z^2>={\nu}<B_\phi^2>
\end{equation}
so comparing this with equation (\ref{10}) we find
$s=(2\nu -1)/\nu$, so
\begin{equation} \label{42}
\nu=1/(2-s).
\end{equation}

Thus for each height we have an $s$ and hence a $\nu$ for that height
and the associated profile is given by solving equation (\ref{38})
with $f=0$ at $\lambda=0$ and $\lambda=1$ and the value of $C$ is
determined by the condition that $f$ is one at its maximum. Notice
that by this means we have determined the weak variation of the
profile with height (see Figure 3).

\subsection{PROFILE FOR CONSTANT EXTERNAL PRESSURE}

From equations (\ref{18}) and (\ref{39}) constant p corresponds to
$\nu=1/2$, $\alpha=0$, a case of great simplicity. Integrating
equation (38) with the boundary conditions that $f=0$ at zero and one,
we find $f'=-C^2(1/2) $ln$(\lambda/\lambda_1)$ and
$f=-C^2(1/2)\lambda$ln$\lambda$ with $\lambda_1=1/e$. The requirement
that $f=1$ at maximum then gives $C^2=2e$ so
\begin{equation} \label{43}
f=-e\lambda ln\lambda.
\end{equation}
With $f$ known we may now evaluate the dimensionless integrals $I$ and$J$,

$I^2=<B_z^2>/\overline{|B_z|}^2=(1/4)\int_0^1(df/d\lambda)^2d\lambda=e^2/4$,
hence
\begin{equation} \label{44}
I=e/2=1.359.
\end{equation}
Likewise
\begin{equation} \label{45}
J=2\sqrt{\int_0^1(f/\lambda)
d\lambda\over\int_0^1(f^{1/2}/\lambda)d\lambda}=\sqrt{2/\pi}=0.798.
\end{equation}
Thus once we specify $\Omega(P)$ and $p$, our solution for the shape
of the magnetic cavity $R_m(z)$ is known via equations (\ref{21}),
(\ref{22}), and (\ref{23}), and within that cavity the structure of
the field is given via the poloidal and toroidal flux
functions $P$ and $\beta$.
\begin{equation} \label{46}
P=P_m(z)e\lambda
\ln(1/\lambda)
\end{equation}

\begin{equation} \label{47}
\beta(P)=4(2\pi^3p)^{1/4}P^{1/2}
\end{equation}
The magnetic fields are given by
$B_R=-(2\pi R)^{-1}\partial P/\partial z, B_\phi = (2\pi R)^{-1}\beta,
B_z=(2\pi R)^{-1}\partial P/\partial R$. These are surprisingly
interesting,
\begin{equation} \label{48}
B_z={1\over\pi R_m^2}{\partial P\over \partial \lambda}={eP_m(z)\over \pi
R_m^2}[\ln(1/\lambda)-1],
\end{equation}
which is positive for $\lambda<1/e$, zero at $\lambda=1/e$, and
negative for $\lambda>1/e$ reaching the value $-eP_m/(\pi R_m^2)$ at
the boundary $\lambda=1$. By equation (\ref{18}) the magnetic pressure
precisely balances the external pressure there. Unexpectedly we find
that $B_z$ is infinite on the axis
\begin{equation} \label{49}
B_z\propto \ln(R_m/R),
\end{equation}
however the flux near the axis is small because $P$ behaves
like $R^2\ln(R_m/R) $ there. Likewise the contribution to the energy from the
magnetic energy-density near the axis behaves as $R^2[\ln(R_m/R)]^2$so the
infinity in the magnetic field appears to be harmless. A greater surprise
comes from the behaviour of $B_\phi$ near the axis.
\begin{equation}  \label{50}
B_\phi={2eP_m\over\pi R_m^2}\sqrt{\ln(R_m/R)},
\end{equation}
whereas we expected to find $B_\phi$ to be zero on axis, it is actually
infinite! However the ratio 
\begin{equation} \label{51}
B_\phi/B_z\rightarrow [\ln(R_m^2/R^2)]^{-1/2}\rightarrow 0 
\end{equation}
thus although the field lines near the axis do wind around it
helically, the helix gets more and more elongated as the axis is
approached so the axis itself is a line of force. Faced with this
example it is evident that the normal boundary condition that on axis
$B_\phi$ should be zero should be replaced by the condition that
$B_\phi/B_z$ should be zero on axis. Since the field is force-free the
currents flow along the field lines and $4\pi{\bf j}=\beta'(P){\bf
B}=(C_1/2)P^{-1/2}{\bf B}$. As $P$ is zero on axis ${\bf j}$ is even
more singular on axis than the magnetic field. Nevertheless the total
current parallel to the axis and crossing any small area is finite and
tends to zero as the area shrinks onto the axis. The exact Dunce's Cap
model of section 6 gives the same field structure close to the
axis. In the next section we find that the infinite fields on axis are
replaced by large finite ones when the pressure decreases at greater
heights. Current sheets are of course a common feature of idealised
MHD and we have one of necessity on the boundary of the magnetic
cavity. The infinite field strengths and current densities on axis
encountered in this solution suggest that very large current densities
occur close to the axis in reality. The lack of sufficient charge
carriers there will lead to a breakdown of the perfect conductivity
approximation with large EMFs appearing up the axis resulting in
particle acceleration along the axis. What observers `see' as a jet
may be just this very high current-density region where the particles
are accelerated, i.e. only the central column of the whole magnetic
cavity which may be considerably wider, cf the observations of
Herculis A, Gizani \& Leahy (2003).

\subsection{PROFILES WHEN PRESSURE DECREASES WITH HEIGHT}
Even when the pressure varies we know how to calculate the shape of
the magnetic cavity, so all we need is the profile of the field across
the cavity.  So we need to solve equation (\ref{38}). For $\nu=1/2$
equation (\ref{43}) gives the solution, but we need it for more
general $\nu$. While this is easily computed an analytical
approximation is more useful. For $\nu<1$ a good approximation is
given by noting that for $\alpha=2\nu-1 $ small $-\ln \lambda
\approx(1-\lambda^\alpha)/\alpha$; using this in our $\alpha=0$
solution for $f$ suggests the form $f\propto
\lambda(1-\lambda^\alpha)/\alpha$ but a better approximation is given
by $f=g(\lambda)/g(\lambda_1)$ where
\begin{equation} \label{52}
g(\lambda)=\alpha^{-1}\lambda(1-\lambda^\alpha)/(1+ a_2\lambda^\alpha),
\end{equation}
and $\lambda_1$ is given by $g'(\lambda_1)=0$. Notice that the
denominator in $g$ is constant when $\alpha=0$ so it makes no
difference to that solution. This form for $f$ automatically satisfies
both the boundary conditions and the one-at-maximum condition. The
equation that gives $\lambda_1$ is a quadratic in $\lambda_1^{\alpha}$
\begin{equation} \label{53}
1-[\alpha+1-a_2(1-\alpha)]\lambda_1^\alpha-
a_2\lambda_1^{2\alpha}=0.
\end{equation}
We use it to find $a_2$ in terms of $\lambda_1$ which
we determine below  
\begin{equation} \label{54}
a_2=[1-\lambda_1^{\alpha}(1+\alpha)]/[\lambda_1^{\alpha}
[1-(1-\alpha)\lambda_1^{\alpha}]].
\end{equation}

The logarithmic infinities in the fields on axis found in equations
(\ref{48})-(\ref{50}) occur because of the log term in equation
(46). The form of equation (\ref{52}) shows that large finite fields
replace those infinities when $\alpha>0$ and so $B_\phi\rightarrow 0$
on axis.

For $\alpha=1$ there is another exact solution in terms of the Bessel
function $J_1$,
\begin{equation} \label{55}
f=k_1\sqrt\lambda
J_1(k_1\sqrt\lambda)/[k_0J_1(k_0)];
\end{equation} 
here $k_0=2.405$ is the first zero of the Bessel function $J_0$ and
$k_1=3.832$ is the first zero of $J_1$. The value of $\lambda_1$ is
therefore $[2.405/3.832]^2=0.3939$. This is not too far from the value
1/e=0.3679 obtained for our $\alpha=0$ solution. This suggests that
linear interpolation i.e.  $\lambda_1= e^{-1}(1+0.07073\alpha)$ and
$a_2$ determined via equation (\ref{53}) will give a good fit and
indeed numerical computations show the fit is excellent all the way
from 0 to 1,the Bessel case.  In the latter the expression
$k_0J_1(k_0)=1.249$.

The integrals $I$ and $J$ can be evaluated for the Bessel solutions
and we give their values for the $\alpha=0$ case in brackets for
comparison.  $I=1.179(1.359);J=1.098(0.799)$. As expected $I$ and $J$
do change with $\alpha$ but remain within 15\% of their means.

Solutions to equation (\ref{38}) for large $\alpha =2\nu-1$ were used
in paper I to solve a different problem but that method works well and
can be extended to work for all $\alpha$ above unity.  Approximate
solutions for large $\nu$ of the form $f=1-\nu^{-1}\ln(\cosh\Lambda)$
where $\Lambda=2\Lambda_1(\lambda-\lambda_1) -\Delta \ln(\cosh[\nu
C(\lambda-\lambda_1)])$ with $\Delta=(6\Lambda_1)^{-1}$;
$\Lambda_1=ch^{-1}e^\nu\simeq\nu$; $C\simeq 2\lambda_1^{1/2}$;
$\lambda_1 \simeq \frac{1}{2}$ are deduced in the appendix as well as
the generalised form
\begin{equation} \label{56}
f=1-H^{-1}\ln(\cosh\Lambda);....;\Lambda= C_*(\lambda-\lambda_1)
-\Delta \ln(\cosh\Lambda),
\end{equation}
with $H=\nu-1/42-5/26.\nu^{-1};~h=\sqrt{H\nu};~\Delta
=1/2.\nu^{-1}-1/9.\nu^{-2}+1/12.  \nu^{-3};$
$\lambda_1=1/2-7/31.\nu^{-1}+4/41.\nu^{-2}$ and
$C_*=hC\lambda_1^{-1/2}=2\nu+\ln4-4/15.\nu^{-1}$. Notice that with
$\Delta $ small equation (\ref{56}) with $H=\nu$ $C_*=\nu$,
$C\lambda^{-1/2}_1$ reduces to our former solution. Inclusion of the
$\Delta$ term allows for the asymmetry around the maximum caused by
the variation of the initial $\lambda$ in equation (\ref{38}). These
freedoms allow us to fit exactly, not only the curvature at the
maximum and the boundary conditions, but also the gradients at which
the solution reaches zero at the ends of the range. Some details are
given in the Appendix. The solutions for $f$ are plotted in Figure 5.
\begin{figure}
\begin{center}
\includegraphics[scale=1.0,angle=0.0,clip]{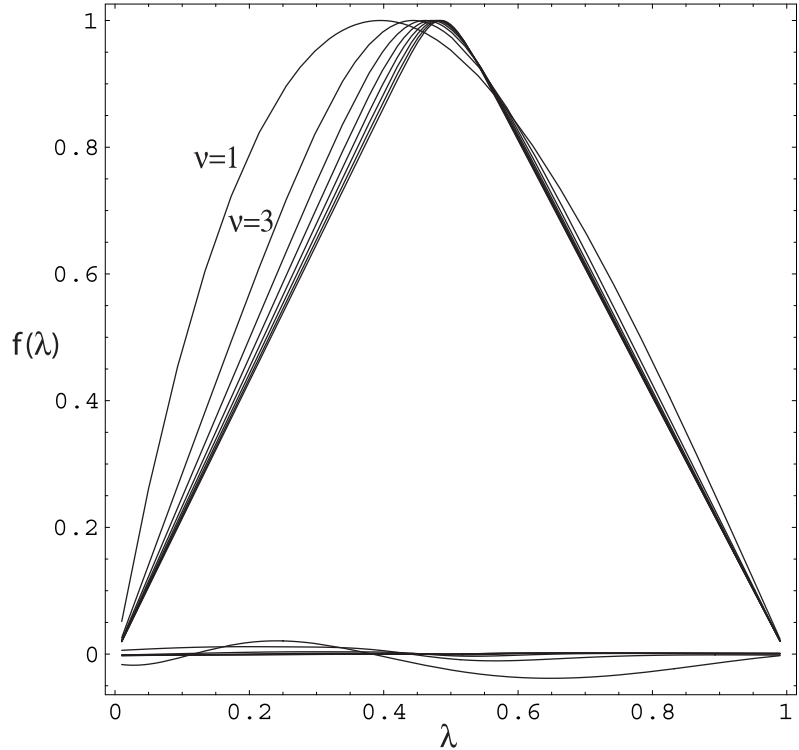}
\caption{Computed profiles $f(\lambda)$ are plotted for
$\nu=1,3,5,...$ . A triangle is the limit $\nu\rightarrow\infty$. At
the bottom are plotted the errors in our analytic approximate
solutions derived in the appendix. Only for $\nu=1$ do the errors rise
above 2\% and for $\nu=1$ itself we have the exact Bessel solution.}
\end{center}
\vspace{-0.2cm}
\label{fig5}
\end{figure}
By taking $\nu$ to be given by equation (\ref{42}) in terms of
$s(z)$ we now have a solution of the form $P=P_mf(\lambda,\nu)$ with
$\lambda=(R/R_m)^2$ and $P_m, R_m$ and $\nu$ all depending weakly on
$z$, see equations (\ref{18}), (\ref{21}) and (\ref{42}). With $P$ so
determined the function $\beta(P)$ is found from $\Phi(P)$ by
integrating along field lines. These are given by ${dR\over
B_R}={Rd\phi\over \\B_\phi}={dz\over B_z}$. Hence
\begin{equation} \label{57}
{dR\over-\partial P/\partial
z}={Rd\phi\over\beta(P)}={dz\over\partial P/\partial R}
\end{equation}
The equality of the first and last terms merely tells us
that $P$ is constant along field lines. The equality of the second and last
terms gives us on integration with $P$ held fixed
\begin{equation} \label{58}
\beta(P)={2P\Phi(P)\over\int (d\ln f/d\ln\lambda)^{-1}dz},
\end{equation}
where the integration is along the curve $P$ constant from one foot
point to the other. While the above procedure is the one to use when
$P(R,0),\Phi(P)$ and $p(z)$ are specified there are special cases that
are simpler. If we ask that $f$ be strictly independent of $z$ rather
than that being an approximation, then equation (\ref{39}) must hold
exactly. Combining that with equation (\ref{18}) we would have
$p=I^2(2\pi)^{-1}(\pi C_2A^{-\alpha})^{-2/(1-\alpha)}\propto
P_m^{2\alpha}$; from (11) this formula leads directly to
$s=2\alpha/(\alpha+1)$ and equation (\ref{37}) gives us the simple
relationship $\beta(P)=C_1P^{(\alpha+1)/2}$ in place of equation (\ref{58}),
nevertheless these equations with s constant imply that $p$ vanishes,
or becomes infinite, when the area $A$ vanishes, which will not be
true. The exactly separable case is too restrictive near the top when
the pressure varies.

\section{Electric Fields, Pushers, Floaters and Squirmers}
We now determine the electric fields that occur when the accretion disk is in
differential rotation. The velocity ${\bf u}$ of a magnetic field line is 
$c{\bf E\times B/B^2}$ and there is no ${\bf E}$ along ${\bf B}$ because of
the perfect conductivity, hence ${\bf E=-u\times B}/c$ . To use this formula
we first calculate the velocity of our field lines. On the accretion disk
this velocity is that of the disk itself so the field line which initially
intersected the disk  at azimuth $\psi$ now has its outer intersection at
$R_o(P)$ at azimuth $\phi_d=\psi+\Omega_d(R_o)t$. We now look for the angular
velocity of the point at which this field line intersects a $z=$const plane.
At each instant the field line obeys equation (\ref{57}) so on integrating the
second equality there
\begin{equation} \label{59}
\phi=\psi+\Omega_d(R_o)t+\Omega(P)tq(z.P)
\end{equation}
where we have written $\Omega(P)t$ for $\Phi(P)$ and $q=\int_0^z(d\ln
f/d\ln\lambda)^{-1}dz/\int(d\ln f/d\ln\lambda )^{-1}dz$. Both
integrations $dz$ are to be performed with $\lambda $ varying so as to
keep $P$ constant.  The final integral is to be performed from
foot-point to foot-point. The physical meaning of q is the fraction of
the total twist on the field line $P$ that occurs by height z. Since
the field line reaches a maximum height $Z(P)$ and then returns, $q$
will be double valued as a function of $z$. To avoid this it is better
to convert those $z-$integrations into integrations over $\lambda $,
in which $q$ is single valued, rather than $z$. Such a conversion
yields
$q=\int_\lambda^{\lambda_o}Z'(P/f)(f\lambda)^{-1}d\lambda/\int_{\lambda_i}^{\lambda_o}
Z'(P/f)(f\lambda)^{-1}d\lambda$, where we have written $P/f$ for $P_m$
to demonstrate that the $\lambda$ dependence arises through $f$ when
$P$ is held fixed. The angular rotation rate of the intersection of
this field line with a $z=$const plane is 
\begin{equation} \label{60}
\dot \phi=\Omega_d+\Omega(P)q+\Omega(P)t\dot q
\end{equation}
$\dot q$ arises from two causes i) any explicit dependence of $Z'$ on
$t$ that does not cancel between numerator and denominator in $q$, ii)
the change in the lower end point $\lambda$ of the numerator's
integration. At constant $P$ and $z$ we have $\partial\ln P_m/\partial
t=-(d\ln f/d\lambda)\dot\lambda$, which gives us the rate of change of
the lower limit of the numerator. When $P(z)$ is either a power or
constant the explicit dependence of $Z'$ on $t$ cancels between
numerator and denominator so there is then no contribution from i).
Of course had we left $q$ as a $z$-integration there would have been
no contribution from the end point but then the contribution from the
$t$-dependence of the $\lambda$ in the integrand must be re-expressed
as a function of $P$ and $z$ via $f(\lambda)=P/P_m(z,t)$, is unwieldy.
Now $R\dot \phi$ at constant $z$ is not the $\phi$-component of the
velocity of the line of force but the velocity of the point of
intersection of the line with the $z$=const plane. When the line of
force is only slightly inclined to the plane the difference is obvious
since the velocity of the line is perpendicular to the line. A little
thought shows that in the intersection velocity the component of the
velocity of the line in the $z$-plane parallel to the projection of
the field into the plane is exaggerated by the factor $B^2/B_z^2$
relative to that component of the field line's velocity. Thus writing
hats for unit vectors and $\hat {\bf b}$ for the unit vector
$(B_R\hat{\bf R}+B_\phi \hat {\bmath \phi})/\sqrt{B_R^2+B_\phi^2}$ in the
$z$-plane considered and $ {\bf u}_\perp$ for the component of the
field line's velocity in that plane, the velocity of the intersection
is ${\bf u}_\perp+[(B^2/B_z^2)-1]({\bf u_\perp.\hat b)\hat b}$. The
component of this intersection's velocity along ${\bf \hat \phi}$ is
\begin{equation} \label{61}
R\dot\phi = u_\phi(1+B_\phi^2/B_z^2)+u_R(B_RB_\phi/B_z^2).
\end{equation}
Now by Faraday's law, the rate of change of the magnetic flux through
a circle about the axis in the plane considered gives the EMF so
\begin{equation} \label{62}
2\pi RE_\phi= -\partial P/\partial ct=-\dot P/c=-(u_zB_R-u_RB_z)/c
\end{equation}
This is a second equation for ${\bf u}$ and the third is just ${\bf
u.B}=0$. Eliminating first $u_z$ and then $u_R$ we find
$u_\phi=[(B_R^2+B_z^2)/B^2]R\dot\phi+[B_RB_\phi/(B^2B_z)]\dot P$. The
other components are readily found from equations (\ref{61}) \&
(\ref{62}). The remaining components of ${\bf E}$ follow from $c{\bf
E=B\times u}$. Like the electric fields on the accretion disk these
fields are not far from cylindrically radial and directed away from
the surface of greatest flux, $P$, at each height.

We now turn our attention to categorising the different types of
solutions.  A Helium balloon needs a tether in tension if it is not to
rise further. A water tank needs a support in compression if it is not
to fall. If we cut a magnetic tower by a horizontal plane there is net
tension across that plane due to the magnetic stresses if
$\int(B_z^2-B_R^2-B_\phi^2)dA/(8\pi) > 0$ and net compression if that
quantity is negative. We call these floaters or pushers respectively
if they satisfy those criteria low down the tower.  Floaters are are
held down by magnetic tension; pushers are supported from the bottom
by a net magnetic pressure. In the tall tower approximation the above
criterion simplifies to $(<B_z^2>-<B_\phi^2>) >0$ for a floater.
Comparing this criterion with equations (\ref{10}) and (\ref{42}) we
see floaters have $s>1,\nu>1,\alpha>1$ and cavities whose radii
increase at greater heights while pushers have $0<s\le 1$, $
0.5<\nu\le 1$ and $0<\alpha\le 1$. Their cavities' radii decrease with
height. In a constant pressure medium all the solutions are
pushers. If the pressure in a long narrow column supporting a weight
is greater than the ambient pressure in the medium surrounding the
column then any lateral bend bowing the column will be exaggerated as
in the Euler strut problem. All our magnetic towers are stable against
that bowing as their net support is atmospheric or less
$(<B_R^2>+<B_\phi^2>-<B_z^2>)/(8\pi)-p=A^{-1}\int A(dp/dz)dz\le
0$. This criterion corresponds to $\alpha\ge 0,s\ge 0$. Thus squirmers
do not occur in the magnetostatics of our systems which have pressure
that decreases with height.

\section{Exact Solutions, Backwards Method}
Here we postulate the forms of $\beta(P)$ and of the bounding surface
S for which we can solve the differential equation (\ref{3}). After we
have solved it we discover what problem we have solved by finding
$p(z)$ on the bounding surface and the twist angles $\Phi(P)$ and the
boundary flux $P(R,0)$. The differential operator on the left of
equation (\ref{3}) separates in oblate or prolate spheroidal and
rotational parabolic coordinates as well as in cylindrical and
spherical polars. It is therefore important to take the surface S to
be one of those coordinate surfaces. Even after choosing one of those
the problem is still non-linear; but there are interesting exceptions.
We have already seen that the case $\beta'\beta =$const. corresponds
to a constant external pressure and it gives a linear but
inhomogeneous equation. A second simple case is $\beta(P)\propto
P\propto \beta'\beta $. This linear case is well known and much
explored but is somewhat less interesting than the $\beta \propto
P^{1/2}$ case. The combination $\beta'\beta \propto P+P_0$ though
soluble has the idiosyncrasies of both the former cases without
obvious advantages. A different approach with more interesting
solutions is to accept the non-linear behaviour of equation (\ref{3})
and look for self-similar solutions.  Most of the above methods can be
generalised using perturbation theory. For example if the bounding
surface is a `cylinder' of slowly varying radius $R_m(z)$ one can
solve the problem in terms of the scaled radius $R/R_m$ as in section
4.

\subsection{SELF-SIMILAR SOLUTIONS}
Writing equation (\ref{3}) in spherical polar coordinates $(r,\theta,\phi)$
and setting $\mu=\cos\theta $ we find
\begin{equation} \label{63}
r^2\partial^2P/\partial
r^2+(1-\mu^2)\partial^2P/\partial \mu^2=-r^2\beta'\beta(P).
\end{equation} 
Self-similar solutions only occur when $P\propto r^{-l},\,\,
\beta(P)=C_1 P^\nu$.

If we now set $P=r^{-l}M(\mu )$ then the left hand side is $r^{-l}$
times a function of $\mu$ so $\beta '\beta $ which is a function of
such a product, must equal such a product. Hence $\beta'\beta
=(1+1/l)C_1^2 P^{1+2/l}$. Thus with $\nu=1+1/l$ the $r^{-l}$ cancels
out and writing $f=M/M_1$, where $M_1$ is the maximum value of
$M$
\begin{equation} \label{64}
l(l+1)f+(1-\mu^2)d^2f/d\mu^2=-(1+1/l) C_3^2
f^{1+2/l}.
\end{equation}
This is a version of the equation of Lynden-Bell and Boily (paper 1)
for whom the case with $l$ small was of especial interest.  \smallskip

To make contact with the work of section 4 we remark that for narrow
jets it is only necessary to consider $\mu$ close to one. Supposing
the walls at $\mu=\mu_m$, we write $\lambda=(1-\mu)/(1-\mu_m)$, where
both $(1-\mu_m)$ and $(1-\mu)$ are small, the equation takes the form
\begin{equation} \label{65}
\lambda d^2f/d\lambda^2+l(l+1)(1-\mu_m)f/2=-C^2(1+1/l)f^{1+2/l}.
\end{equation}
We have approximated $1-\mu^2$ as $2(1-\mu)$, which is good to 1 per
cent or better for opening angles of less than 23 degrees. When the
$(1-\mu_m)l(l+1)f/2$ is neglected this equation will be recognised as
equation (\ref{38}) of section 3. The particular solutions of equation
(\ref{64}) to which we now turn are so simple that we obtain them
without making this narrow cone approximation.

\subsection{THE DUNCE'S CAP MODEL}
Unexpectedly we shall use spherical polar coordinates centred at the
top of the tower-like magnetic cavity at $(0,0,Z)$. This use of $Z$ is
similar to the very crude description as a cylindrical tower in the
introduction but now the tower will be conical pointing down at the
accretion disk. This $Z$ corresponds to $Z_h$ in sections 3 and 4. We
measure $\theta$ from the downward axis pointing toward the centre of
the accretion disk (see Figure 6). We solve the force free equations
within the downward opening cone cos$\theta\ge\mu_m$ which forms a
dunce's cap configuration over the accretion disk. We take $l=-2$
which corresponds to our $\beta\propto P^{1/2}$ behaviour appropriate
for a constant pressure. Equation (\ref{64}) becomes $2f+
(1-\mu^2)d^2f/d\mu^2=-C_4$. Writing $f=(1-\mu^2)m$ our equation
becomes

$d/d\mu[(1-\mu^2)^2dm/d\mu]=-C_4$, so, using the boundary condition that
$(1-\mu^2)m'$ is non-singular at $\mu=1$, we find,  
$m=C_4\int(1+\mu)^{-2}(1-\mu)^{-1}d\mu$. On integration by parts we get
\begin{equation} \label{66}
P=r^2(1-\mu^2)m=C_4r^2(1-\mu^2)\left[{1\over 2}(1+\mu_m)-{1\over
2}(1+\mu)+{1\over 4}\ln\left({1+\mu\over1+\mu_m}.{1-\mu_m\over
1-\mu}\right) \right].
\end{equation}

\begin{figure}
\begin{center}
\includegraphics[scale=0.6,angle=0.0,clip]{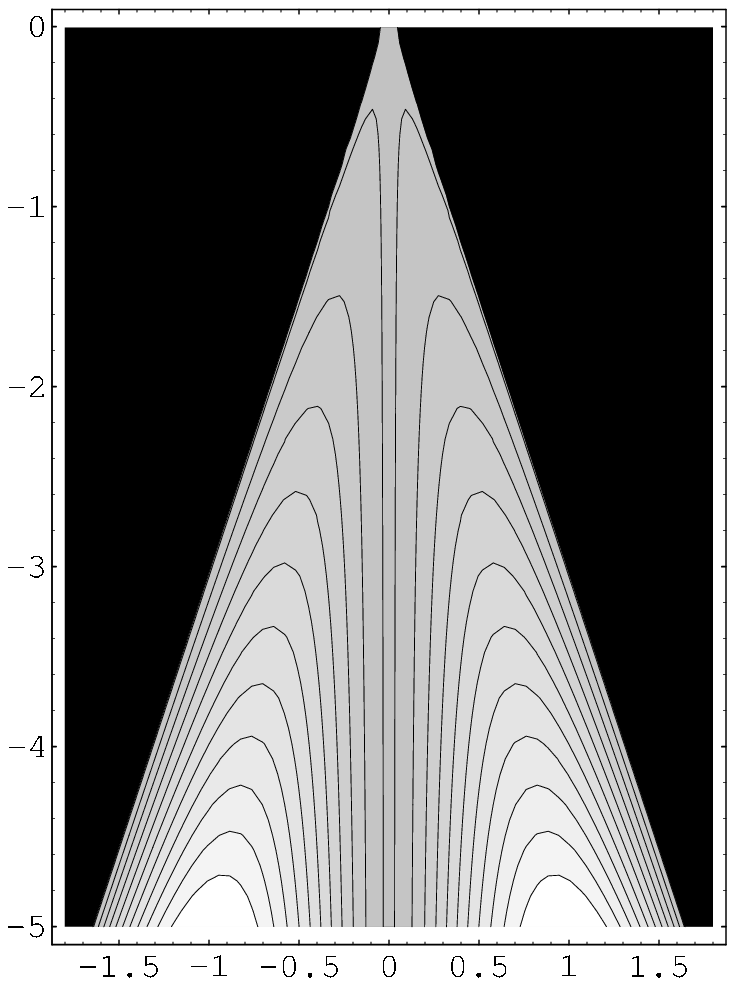}
\includegraphics[scale=0.6,angle=0.0,clip]{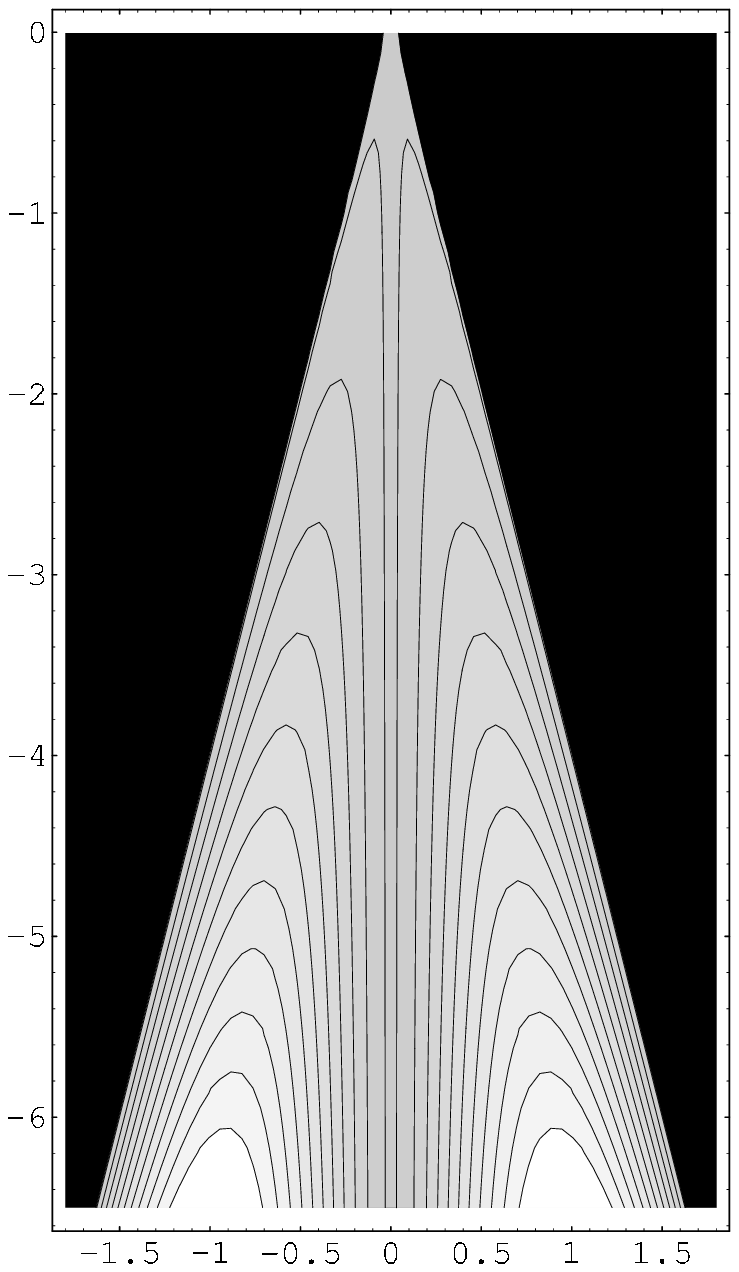}
\caption{The poloidal lines of force in the Dunce's cap model at two
different times. The top advances at constant speed while the bottom only
revolves in differential rotation}
\label{duncecap}
\end{center}
\vspace{-0.2cm}
\label{fig6}
\end{figure}

Evidently $P=0$ on $\mu=\mu_m$ and on the axis. For tall narrow cones $\theta$
is small,$\mu,\mu_m$ are near one and the flux function becomes
$P={1\over4}C_4r^2\theta^2[\ln(\theta_m^2/\theta^2)].$ At $r=Z$ the maximum
value of $P$ is $F$ so $C_4=4eF/R_d^2$ where
$R_d=\theta_mZ.$ Also $\beta(P)=\sqrt{2|C_4|}P^{1/2}=
C_4r\theta[\ln(\theta_m/\theta)]^{1/2}$.
Also, see Figure 7, 
\begin{equation} \label{67}
2\pi B_r=(r^2\theta)^{-1}\partial P/\partial
\theta=C_4[\ln(\theta_m/\theta)-1];\,2\pi B_\theta=
C_4\theta\ln(\theta_m/\theta);\,2\pi B_\phi=
C_4[\ln(\theta_m/\theta)]^{1/2} 
\end{equation}

\begin{figure}
\begin{center}
\includegraphics[scale=1.0,angle=0.0,clip]{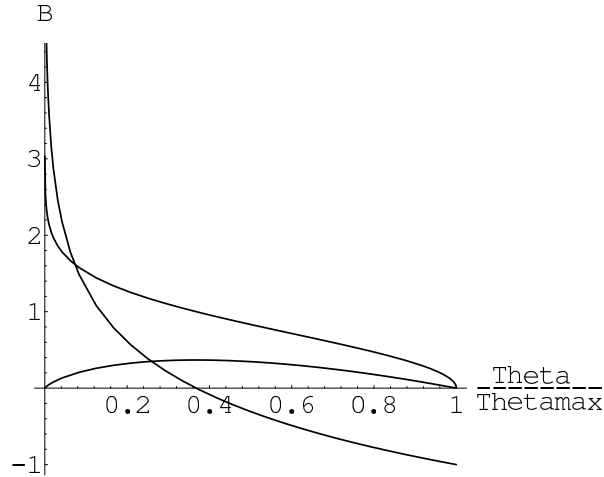}
\caption{$B_r,B_\theta$ and $B_\phi$ are all independent of $r$ inside the
Dunce's Cap so we plot them as functions of $\theta/\theta_{max}$. While
$B_\theta$ is zero at both boundaries $ B_\phi$ is zero only at the outer one.
$B_r$ passes through zero at an internal point. See text for the surprising
behaviour near $\theta=0$}
\label{Bvstheta}
\end{center}
\vspace{-0.2cm}
\label{fig7}
\end{figure}

The constant $C_4$ is determined by the external pressure $C_4 =
8\pi[\pi p]^{1/2}$. Near the axis these fields have precisely the same
behaviour as those we found in equations (\ref{49}) and
(\ref{50}). However the fields are in one sense even more interesting;
they are all independent of $r$, the distance from the origin. By
construction $B_\theta,B_\phi $ vanish at the edge of the cone but
$B_r$ is constant along its generators all the way up to the
origin. The origin itself is then a point at which the infinite field
coming up the axis turns back and splays out down these generators at
constant field strength.  This is the way the constant external
pressure is opposed.  With the origin such a remarkable point, some
may suspect that the forces do not balance there, that the whole
configuration might be held up by some sky-hook pulling at the
apex. However $P$ varies as $r^2$ and the Maxwell stresses integrated
over horizontal cuts through the cone give a net force that vanishes
like $r^2$ as the origin is approached. Thus no sky hook is necessary
to balance forces at the origin. The radial fields reverse at an
intermediate value $\theta_1=\theta_m/\sqrt e$ (for narrow cones) and
some may suspect reconnection there but the $B_\phi$ field component
maximises at that value of theta so there is plenty of magnetic
pressure to prevent reconnection. With $\mu_m$ close to one the radii
on the accretion disk are $R=\theta Z$ so the flux coming up within a
circle of radius $R$ is $P(R,0)=Fe(R^2/R_d^2)[\ln(R_d^2/R^2)]$. The
twist function $\Phi(P)$ is given by integration along field lines
which is simple for this self-similar field.
$rd\theta/B_\theta=r\sin\theta d\phi/B_\phi$ yields on writing
$\eta=\theta_m/\theta$ 
\begin{equation} \label{68}
\Phi(P)=\Omega(P)t=(Z/R_d)\int_{\eta_i}^{\eta_o}(\ln\eta)^{-1/2}d\eta\simeq
(Z/R_d)[(\eta_i-\eta_o)(\ln\eta_i)^{-1/2}],
\end{equation}
where the integral is evaluated between foot points at $R_{i,o}(P)$ of
the field line and $\eta_{o,i}(P)=R_d/R_{o,i}(P)$. Therefore
$\eta_{o,i}$ are the roots for $\eta$ of
\begin{equation} \label{69}
P/F=2e\eta^{-2}\ln \eta
\end{equation}
When the inner foot is much further in than the outer foot then
$\Omega_i>>\Omega_o$ so then $\Omega(P)$ is close to
$\Omega_i(P)$. This allows us to estimate the angular velocity of the
accretion disk required. In the central parts it behaves as
$R^{-1}[\ln(R_d/R)]^{-1/2}$, so the circular velocity of the disk is
almost constant but drops to zero at the very centre like
$[\ln(R_d/R)]^{-1/2}$. From Equation (\ref{69}) both roots
$\eta_{o,i}$ are independent of time. Hence from equation (\ref{68})
$Z\propto t$ as expected from section 4 so we may write
$Z=Vt$. Writing our solution for P in terms of cylindrical polar
coordinates now centred on the centre of the disk,
$P=eF(R^2/R_d^2)\ln(R_m^2/R^2)$, where $R_m(z)=R_d(1-z/Z)$ for $z<\,Z$
and it is just the dependence of $Z$ on time that generates the
evolution of the system. $\beta(P)=\sqrt{8eF}P^{1/2}/R_d$ with
$\Phi(P)$ given by equation (\ref{69}). The magnetic fields are given by
equation (\ref{67}) so we now calculate the electric fields following
the method of section 5. The simplicity of the expression for $P$
means that $q$ is simpler and in place of equation (\ref{60}) we have
$\phi=\psi+\Omega_d(R_o)t+(Vt/R_d)\int_{\eta_o}^{R_m/R}(\ln\eta)^{-1/2}d\eta
$\,. Now $R_m= (1-{z \over Vt})R_d$ so $\dot R_m=zR_d/(Vt^2)4$. At
fixed height a point only remains on the same field line if $\dot
R=-\dot P/(\partial P/\partial R)=-\dot P/(2\pi RB_z)$ Using these and
differentiating $\phi$ at constant $z$ and $P$ 
\begin{equation} \label{70}
\dot \phi=\Omega_d+VR_d^{-1}\int_{\eta_o}^{R_m/R}(\ln \eta)^{-1/2}d
\eta+[\ln(R_m/R)]^{-1/2}\left({z\over Rt}+{\dot PVt\over 2\pi
R_dR^3B_z}\right).
\end{equation}
As before Equations (\ref{61}) and (\ref{62}) give the velocities of
the lines of force
as,
\begin{eqnarray} \label{71}
u_\phi&=&[(B_R^2+B_z^2)/B^2]R\dot \phi
+[B_RB_\phi/(B^2B_z)]\dot P\nonumber \\ u_R &=&
[B_z^2/(B_RB_\phi)][R\dot \phi-u_\phi(1+B_\phi^2/B_z^2)]\\ u_z &=&
(B_z/B_R)u_R+\dot P/B_R \nonumber
\end{eqnarray}
Also $cE_\phi=-\dot P/(2\pi R)$ while the other components follow from
$c{\bf E=B\times u}$. Notice that the meridional field velocity ${\bf
u}_M=(u_R,u_z)$ is not perpendicular to the vector meridional
component of the magnetic field.  It is the full vector velocity that
is perpendicular to the complete magnetic field.  We have explored the
Dunce's Cap model in all this detail both to demonstrate its
interesting field structure and to show how we get the fields in an
explicit example. Figure 5 gives the poloidal structure of the
magnetic fields which wind around these surfaces of constant flux.

\section{Conclusions}
It is very remarkable that these simple analytic methods allow the
solution of these highly non-linear problems with variable domain, not
just for a few special cases but for all twist functions Omega and all
pressure distributions. Furthermore we get time-dependent solutions
for all time.  The main secret lies in the variational principle
coupled with a good form of trial function. Once the time-dependence
becomes relativistic we lose this tool so the problems will become
harder. However when the relativistic motion is only important for the
rotation it is likely that another variational principle may exist. An
important problem is to seek it out.  In the foregoing we have
explored the simplest case in which the magnetic cavity is empty and
the Poynting flux carries both the energy and the momentum of the
jet. I believe that the results justify the assertion that this
simplest case has remarkable similarities to the observed world and
this suggests that the extra complication of material winds is not
essential to explaining the main phenomena. However especially for
Pulsars Relativistic rotation is an essential ingredient of the
problem not covered here.

\section{Acknowledgments}
We are grateful to the Institute of Advanced Study for providing the
environment at Princeton where much of this work was done thanks to
the support of the Monell Foundation. It was started earlier on a
visit to the Carnegie Observatories in Pasadena where most of paper
III was completed.  Discussions with N.O.Weiss and P.Goldreich proved
most helpful; C.Pichon and G.Preston helped with computations and
F.J. Dyson was a source of encouragement.

\appendix

\section[]{A useful mathematical technique}
We wish to solve a highly non-linear differential equation such as
\begin{equation} \label{A1}
(1-\mu^2) \frac{d^2f}{d\mu^2}=-l(l+1)f-C^2\nu
f^{2\nu-1}~;~\nu=1+\frac{1}{l}
\end{equation}
under the boundary conditions that $f$ is zero at given end points and
that $C$ is so chosen that the maximum of $f$ is one. There should be
only only maximum of $f$ between the end points. We rewrite the
equation in the form
\begin{equation} \label{A2}
(1-\mu^2)\frac{d}{d\mu}\left(\frac{df}{d\mu}\right)^2 =\frac{d}{d\mu}S^2
\end{equation}
where
\begin{equation} \label{A3}
S^2=l(l+1)(1-f^2)+C^2(1-f^{2\nu})~.
\end{equation}
We shall be particularly interested in the strongly non-linear case
with $\nu$ large ($l$ small) so we start with $\nu$ large but later
extend the technique down to $\nu\approx 1$. As we shall come across
several variants of essentially the same problem we describe the
technique in terms of the more general problem in which \ref{A2} is
replaced by 
\begin{equation} \label{A4}
G(\mu) \frac{d}{d\mu}\left(\frac{df}{d\mu}\right)^2 = \frac{d}{d\mu}
[S(f)]^2
\end{equation}
with $S^2(f)$ zero when $f$ reaches its maximum of 1. We shall assume
that $S(f)$ depends on parameters such at $\nu$ or $C$ in (\ref{A1}) and
that we are primarily interested in the solution when $\nu$ is large
in which case $dS^2/df$ is strongly peaked close to $f=1$ and away
from there it is small. Under these conditions (which hold for
(\ref{A1})) most of the deceleration $-d^2f/d\mu^2$ of $f$ occurs in the
region in which $f$ is close to one. Let that maximum be at
$\mu=\mu_1$. Then at $\mu_1$
\begin{equation} \label{A5}
(d^2f/d\mu^2)_1 =\frac{1}{2G_1}\left[\frac{dS^2}{df}\right]_1= -g~{\rm say}
\end{equation}
where $G_1=G(\mu_1)$. Near $\mu=\mu_1$
\begin{equation} \label{A6}
f= 1-\frac{1}{2}g(\mu-\mu_1)^2
\end{equation}
and $S^2=(dS^2/df)_1(f-1)$ so
\begin{equation} \label{A7}
\mu-\mu_1 =G_1^{-1/2} g^{-1}S~
\end{equation}
\begin{equation} \label{A8}
G(\mu)\simeq G_1+G_1^\prime(\mu-\mu_1) \simeq G_1(1+a_1 S)^3
\end{equation}
where
\begin{equation} \label{A9}
a_1=\frac{1}{3}G_1^\prime G_1^{-3/2}g^{-1}
\end{equation}
and $G_1^\prime =(dG/d\mu)$ at $\mu_1$.

\noindent Writing (\ref{A8}) as a cubic leads to simpler mathematics
later. Near $\mu=\mu_1$ where most of the `deceleration' of $f$ takes
place we see from (\ref{A4}) that
\begin{equation} \label{A10}
\frac{d}{d\mu}\left(\frac{df}{d\mu}\right)^2 = \frac{2S
dS/d\mu}{G_1(1+a_1 S)^3}
\end{equation}
We have derived this equation near $\mu=\mu_1$ but away from there $f^{2\nu}$
is small and as $dS^2/d\mu$ is small and $1+a_1S$ is not near zero, the
equation can be taken to hold everywhere. We now integrate (\ref{A10})
and remembering that $df/d\mu$ is zero at $\mu_1$ we find
\begin{equation} \label{A11}
(df/d\mu)^2=\frac{S^2}{G_1(1+a_1S)^2}
\end{equation}
It was to get this simple result that we chose the cubic form in
(\ref{A8}). Taking the square root and multiplying up by $1/S+a_1$ we
find
\begin{equation} \label{A12}
\int^{\mu_1}_\mu \frac{df}{S} +a_1(1-f) = G_1^{-1/2}(\mu-\mu_1)
\end{equation}
To proceed further we must integrate $df/S(f)$. As the major item in
(\ref{A1}) is $f^{2\nu -1}$ we shall assume that near the maximum of
$f$, $S^2$ can be approximated as $C^2_1(1-f^{2\overline{\nu}})$. We
do this by taking
\begin{equation} \label{A13}
\overline{\nu} =\frac{1}{2}[d\ln[S^2(0)-S^2]/d\ln f]_{f=1}
\end{equation}
and
\begin{equation} \label{A14}
C_1^2=gG_1/\overline{\nu}
\end{equation}
We now use the method of paper I to give us $\int
(1-f^{2\overline{\nu}})^{-1/2}df$ for large $\overline{\nu}$.

\noindent Set $f=1-\frac{1}{\overline{\nu}}\ln
q;~df=-\frac{1}{\overline{\nu}}\frac{dq}{q}$.  Then
$f^{2\overline{\nu}}=(1-\frac{1}{2\overline{\nu}}\ln
q^2)^{2\overline{\nu}}\rightarrow \exp -(\ln q^2)=1/q^2$ so\\$\int
(1-f^{2\overline{\nu}})^{-1/2}df =\frac{1}{\overline{\nu}} \int(q^2-1)^{-1/2} dq
=\frac{1}{\overline{\nu}} ch^{-1}q$ so setting $q=ch \Lambda$ we find
\begin{equation} \label{A15}
f=1-\frac{1}{\overline{\nu}}\ln(ch\Lambda )
\end{equation}
and from (\ref{A12}( $\Lambda$ is given by 
\begin{equation} \label{A16}
\Lambda +a_1C_1 \ln (ch\Lambda) =\overline{\nu}C_1G_1^{-1/2}(\mu-\mu_1)
\end{equation}
and we remember that $a_1
=\frac{1}{3}G^\prime_1G_1^{-3/2}g^{-1}=\frac{1}{3}
\frac{G^\prime_1}{G_1^{1/2}\overline{\nu}C_1^2}$. To complete the solution
we must impose the boundary conditions and so find the value of $C$
needed to make $f$ one at its maximum. However the boundaries vary
from one application to another so it is time to consider the
differential equations and their ranges one by one.

The application in this paper is to the solution of equation
(\ref{38}) $\lambda d^2f/d\lambda^2=-C^2\nu f^{2\nu-1}$. 

To use the general theory above for this equation we write $\lambda$
for $\mu$ and $G(\lambda)=\lambda$, $G_1=\lambda_1$ and
$S^2=C^2(1-f^{2\nu})$ the boundaries are at $\lambda=0$ and
$\lambda=1$ and $\overline{\nu}=\nu$, $C=C_1$. At those boundaries we need
$ch\Lambda =e^\nu$ so that $f=0$. If $\Lambda_1$ is the positive
value of $\Lambda$ satisfying this then the negative solution is
$-\Lambda_1$ so at $\lambda=1$ and $\lambda=0$ respectively
$\Lambda_1+a_1C\nu=\nu C\lambda_1^{-1/2}(1-\lambda_1)$

\noindent $-\Lambda_1+a_1 C\nu = -\nu C \lambda_1^{1/2}$ where
$a_1=\frac{1}{3\lambda_1^{1/2}\nu C^2}$.

\noindent Subtracting $2\Lambda_1=\nu C \lambda_1^{-1/2}$

\noindent adding $a_1C\nu=\nu C\lambda_1^{-1/2}(\frac{1}{2}
-\lambda_1)=2\Lambda_1 (\frac{1}{2}-\lambda_1)$ so using the value of
$a_1$ above and eliminating $C$
$$\lambda_1\left(\frac{1}{2}-\lambda_1\right)=\nu/(12\Lambda^2_1)$$

Now $\Lambda_1 \simeq \nu$ or more accurately $\nu\left(1+\frac{\ln
2-\frac{1}{4}e^{-2\nu}}{\nu}\right)$ so the quantity on the right is
$0\left(\frac{1}{12\nu}\right)$ which is small so $\lambda_1$ is nearly
a half. Accurately
\begin{equation} \label{A17}
\lambda_1=\frac{1}{2}\left[\frac{1+\sqrt{1-
\frac{4\nu}{3\Lambda_1^2}}}{2}\right] 
\end{equation}
with $\lambda_1$ known 
\begin{equation} \label{A18}
C=2\lambda_1^{1/2}\Lambda_1/\nu
\end{equation}
This completes the solution when $\nu$ is large
\begin{equation} \label{A19}
f\simeq1-\frac{1}{\nu}\ln\left\{ch\left\{2\Lambda_1(\lambda-\lambda_1)-
\frac{1}{6\Lambda_1}
\{\ln
ch[2\Lambda_1(\lambda-\lambda_1)]\}\right\}\right\}
\end{equation}

\subsection{Solutions when $\nu$ is no longer large}
When $\nu$ is lowered the solution will still have a single hump and
will still fall to zero at the end points but it will no longer have
precisely the form given in (\ref{A15}),(\ref{A16}) because there will be
more deceleration of $f$ away from $\mu=\mu_1$ where our
approximations for $G(\mu)$ and $S(f)$ are no longer valid. To allow
for this we generalise the form of (\ref{A16}) and (\ref{A17}) to
\begin{equation} \label{A20}
f=1-\frac{1}{H} \ln ch\Lambda
\end{equation}
\begin{equation} \label{A21}
\Lambda + \Delta \ln ch \Lambda = h CG_1^{-1/2}(\mu-\mu_1)
\end{equation}
where $H$ and $h$ can now be different and only for $\nu$ large will
they become equal to $\nu$. Furthermore $\Delta$ is no longer
restricted to the form it took previously in terms of $a_1,~\nu$ and
$C$. We show in this section that with the four parameters
$H,~\Delta,~h~C$ and $\mu_1$ we can ensure that the solution has the
right position of the maximum, the right deceleration there and that
the gradients of the solution at the end points are appropriate for
the differential equation. Since the desired solution has just one
hump and is zero at the end points it is no surprise that we can fit
it well not only for $\nu$ large but for $\nu > 1$. In practice we
find errors from computed solutions are less than 1\% rising above 2\%
for $\nu=1$. Since $C$ is itself to be determined from the fitting of
boundary conditions in the original equation we have actually five
constants $H,~\Delta,~h,~C$ and $\mu_1$ to be determined so we impose the
following five conditions on our proposed form of solution A
\begin{equation} \label{A22}
1/~\&~2/~f=0~{\rm at~the~end~points~so~there}~ch\Lambda_1=e^H
\end{equation}
\begin{equation} \label{A23}
3/~~~-\frac{d^2f}{d\mu^2}=-\frac{1}{2G_1}\left(\frac{dS^2}{df}\right)_{f=1}
~~~{\rm
this~gives}~\frac{h^2C^2}{H}=\frac{1}{2}~\left.\frac{dS^2}{df}\right|_1
\end{equation}
As we wish our proposed solution to solve (\ref{A4}) as nearly as
possible we now integrate (\ref{A4}) by parts first from $\mu_1$ to 1 and
then from the lower boundary $\mu_m$ to $\mu_1$. These give us
\begin{equation} \label{A24}
G(1)\left(\frac{df}{d\mu}\right)^2_{\mu=1} -
\int^1_{\mu_1}\left(\frac{df}{d\mu}\right)^2 G^\prime(\mu)d\mu=[S(0)]^2
\end{equation}
and
\begin{equation} \label{A25}
-G(\mu_m)\left(\frac{df}{d\mu}\right)^2_{\mu_m} +
 \int^{\mu_1}_{\mu_m} \left(\frac{df}{d\mu}\right)^2
 G^\prime(\mu)d\mu=[S(0)]^2
\end{equation}
We now use the forms (\ref{A20}) and (\ref{A21}) and substitute
\begin{equation} \label{A26}
\left(\frac{df}{d\mu}\right)^2 =\frac{1}{H} th^2\Lambda
\frac{d\Lambda}{d\mu} \frac{hC}{G_1^{1/2}[1+\Delta th\Lambda]}
\end{equation}
into the integrals in (\ref{A24}),(\ref{A25}) but use for the end value in
(\ref{A25})
\begin{equation} \label{A27}
\left(\frac{df}{d\mu}\right)^2_{\mu_m} =\frac{1}{H^2} (1-e^{-2H})
\frac{h^2C^2}{G_1[1-\Delta\sqrt{1-e^{-2H}}]}
\end{equation}
For $(f^\prime)^2_{\mu=1}$ we merely reverse the sign of $\Delta$ in
(\ref{A27}). We then perform the integrals, so (\ref{A22})--(\ref{A25})
provide us with the five equations to determine our five parameters.

We wrote (\ref{A26}) in that form because in the application to narrow
jets $G$ is linear so that the integrals are readily performed using
$\Lambda$ as the variable of integration. For that case the integral
involved is

$$\int \frac{th^2\Lambda}{1+\Delta th\Lambda}d\Lambda =\int
\frac{1-sech^2\Lambda}{1+\Delta th\Lambda}d\Lambda =-\frac{1}{\Delta}
(1+\Delta th\Lambda) +\int
\frac{(e^{2\Lambda}+1)d\Lambda}{(1+\Delta)e^{2\Lambda}+1-\Delta}$$
$$ =\frac{-1}{\Delta} \ln (1+\Delta th \Lambda)
+\frac{\Lambda}{1-\Delta^2} - \frac{\Delta}{1-\Delta^2}\ln
\left(\frac{1+\Delta}{1-\Delta} +e^{-2\Lambda}\right)$$

Applying these methods to the solution of equation (\ref{38}), with
$\lambda$ written for $\mu$ and $G(\lambda)=\lambda,~S^2=C^2(1-f^{2\nu})$
and boundaries at $\lambda=0$ and $\lambda=1$ the five equations are
with $\Lambda_1=ch^{-1}e^H$
\begin{equation} \label{A29}
\Lambda_1 +\Delta \ln ch \Lambda_1 =hC\lambda_1^{-1/2}(1-\lambda_1)
\end{equation}
\begin{equation} \label{A30}
-\Lambda_1 +\Delta \ln ch\Lambda_1 =-hC\lambda_1^{+1/2}
\end{equation}
\begin{equation} \label{A31}
h^2/H=\nu
\end{equation}
and from (\ref{A24}) and (\ref{A25})
\begin{equation} \label{A32}
C^2= \frac{1}{H^2} (1-e^{-2H})
\frac{h^2C^2}{\Lambda_1(1+\Delta\sqrt{1-e^{-2H}})} -
\frac{hC\lambda_1^{-1/2}}{H^2} \left\{-\frac{1}{\Delta} \ln
\left[1+\Delta\sqrt{1-e^{-2H}}\right]+\frac{\Lambda_1}{1+\Delta}
-\frac{\Delta}{1-\Delta^2} \ln
\left(\frac{1+\sqrt{1-e^{-2H}}}{1+\Delta\sqrt{1-e^{-2H}}}\right)\right\}
\end{equation}
\begin{equation} \label{A33}
C^2= \frac{hC\lambda_1^{-1/2}}{H^2} 
\left\{+
\frac{1}{\Delta}\ln
(1-\Delta\sqrt{1-e^{-2H}}) + 
\frac{\Lambda_1}{1\Delta} -
\frac{\Delta}{1-\Delta^2} \ln
\left(
\frac{1-\sqrt{1-e^{-2H}}}{1-\Delta\sqrt{1-e^{-2H}}}
\right)
\right\}
\end{equation}
In solving these one uses the sum and differences of (\ref{29}) and
(\ref{30}) and thinks of $H$ rather than $\nu$ as the independent
variable. Figure 5 demonstrates the accuracy of the results compared
with computed solutions.

Finally we rewrite the approximate results as power series in
$\nu^{-1}$

\noindent$H=\nu-1/42-5/26.\nu^{-1}$\\ 
$h=\sqrt{H\nu}$\\
$\Delta=1/2.\nu^{-1}-1/9.\nu^{-2}+1/12.  \nu^{-3}$\\
$\lambda_1=1/2-7/31.\nu^{-1}+4/41.\nu^{-2}$\\
$C_*=hC\lambda_1^{-1/2}=2\nu+\ln4-4/15.\nu^{-1}$

\label{lastpage}

\end{document}